\documentclass[acmsmall,screen]{acmart}

\usepackage{color,soul}
\usepackage{pifont}
\usepackage{tablefootnote}
\usepackage{soul}
\usepackage{wrapfig}
\usepackage{adjustbox}
\usepackage{subcaption}

\newcommand{\code}[1]{{\ttfamily \small #1}}

\newcommand{\approachName}{LintQ}
\newcommand{\core}{LintQ-core}
\newcommand{\nPatternDetectors}{ten}
\newcommand{\NPatternDetectors}{Ten}

\newcommand{\KeepOnlyPyAndIpynb}{75,304}

\newcommand{\RemoveBlacklistedRepos}{41,739}

\newcommand{\TotalLinesOriginalPrograms}{757,369}

\newcommand{\avgGateCallsPerFileNoMeasure}{9.9}
\newcommand{\avgMeasureCallsPerFile}{1.2}
\newcommand{\topFirstGateName}{x}
\newcommand{\topSecondGateName}{h}
\newcommand{\topThridGateName}{cx}

\newcommand{\miniSurveyExclusionReasonSpecRequired}{16}
\newcommand{\miniSurveyExclusionReasonNotBugList}{16}
\newcommand{\miniSurveyExclusionReasonNotOnPrograms}{16}
\newcommand{\miniSurveyExclusionReasonHardwareRelated}{13}
\newcommand{\miniSurveyExclusionReasonOnlyMentioningQuantum}{12}
\newcommand{\miniSurveyExclusionReasonOnlyUsingQuantum}{9}
\newcommand{\miniSurveyExclusionReasonFieldOverview}{8}
\newcommand{\miniSurveyIncludedWorks}{five}
\newcommand{\miniSurveyExclusionReasonTheoryPaper}{2}
\newcommand{\miniSurveyExclusionReasonDuplicate}{2}

\newcommand{\miniSurveyBugPatternRequireHardwareKnowledge}{5}
\newcommand{\miniSurveyBugPatternSpecRequired}{5}
\newcommand{\miniSurveyBugPatternRuntimeInfo}{1}
\newcommand{\miniSurveyBugPatternLintqModelingLimitations}{7}
\newcommand{\miniSurveyBugPatternNotAnIssue}{3}
\newcommand{\miniSurveyBugPatternTooApiSpecific}{9}

\newcommand{\miniSurveyTotalBugPatternsNoDuplicates}{39}
\newcommand{\miniSurveyTotalAnalysesImplemented}{ten}
\newcommand{\miniSurveyPatternsFromDevDiscussion}{three}

\newcommand{\avgAbstractionsUsedPerQuery}{3.1}
\newcommand{\avgQueriesSupportedPerAbstraction}{5.5}

\newcommand{\avgLoCQuery}{10}
\newcommand{\minLoCQuery}{3}
\newcommand{\maxLoCQuery}{17}

\newcommand{\nProgramsSelectedQiskit}{7,568}

\newcommand{\dataDatasetMiningQiskit}{Feb 14, 2023}

\newcommand{\percAgreement}{70.1\%}

\newcommand{\inspectionSizePerOurDetector}{ten}
\newcommand{\inspectionSizePerOurDetectorNum}{10}

\newcommand{\nInspectedWarnings}{345}
\newcommand{\nTruePositivesLintQ}{216}
\newcommand{\allTPOverAllWarningsLintQ}{62.6\%}
\newcommand{\medianPrecisionLintQ}{69.3\%}

\newcommand{\nCheckersAboveFiftyPrecision}{six}
\newcommand{\nInspectedWarningsBestCheckers}{133}
\newcommand{\nTruePositivesLintQBestCheckers}{121}
\newcommand{\allTPOverAllWarningsLintQBestCheckers}{91.0\%}

\newcommand{\nConfirmed}{seven}

\newcommand{\TPReported}{70}
\newcommand{\TPUnreportedBecauseTeachingCode}{47}
\newcommand{\TPUnreportedBecauseUnmantained}{31}
\newcommand{\TPUnreportedBecauseTestingOtherFeatures}{27}
\newcommand{\TPUnreportedBecauseIncompletePersonalProgramSketches}{16}
\newcommand{\TPUnreportedBecauseIntermediateVisualization}{16}
\newcommand{\TPUnreportedBecauseArchived}{9}

\newcommand{\nWarningsOnLintQTPwithQChecker}{200}
\newcommand{\nWarningsOnLintQTPwithQCheckerOverlap}{9}
\newcommand{\nWarningsOnLintQTPwithQCheckerMissed}{207}
\newcommand{\nWarningsOnLintQTPwithQSmell}{77}
\newcommand{\nWarningsOnLintQTPwithQSmellOverlap}{0}

\newcommand{\nWarningsOnLintQTPwithPylint}{8,627}
\newcommand{\nWarningsOnLintQTPwithPylintOverlap}{42}
\newcommand{\nWarningsOnLintQTPwithPylintMissed}{174}
\newcommand{\nTPsMissedByAllCompetitors}{199}
\newcommand{\percTPsMissedByAllCompetitors}{92.1\%}
\newcommand{\nTPsFoundByAtLeastOneCompetitor}{17}
\newcommand{\percTPsFoundByAtLeastOneCompetitor}{7.9\%}

\newcommand{\totalDatasetQueryCompilationTimeSec}{97.0}

\newcommand{\totalDatasetEvaluationTimeMin}{162}

\newcommand{\totalDatasetCreationTimeMin}{74}
\newcommand{\avgPerProgramEvaluationTimeSec}{1.3}

\newcommand{\topThreeDetectorRuleOne}{DoubleMeasurement}
\newcommand{\topThreeDetectorRuleOneTimeSec}{2,290}
\newcommand{\topThreeDetectorRuleTwo}{OpAfterMeasurement}
\newcommand{\topThreeDetectorRuleTwoTimeSec}{1,761}

\newcommand{\nInspectedWarningsQSmellStatic}{20}

\newcommand{\nDetectorsWithWarningsQSmellStatic}{two}

\newcommand{\nInspectedWarningsQChecker}{59}

\newcommand{\nDetectorsWithWarningsQChecker}{six}
\newcommand{\nQcheckerTP}{three}

\newcommand{\recallBugsFourQ}{7.1\%}

\usepackage{xfp} %
\usepackage{color}
\newcommand{\barwidth}{4} %
\newcommand{\barheight}{4pt} %

\definecolor{salmon}{rgb}{1.0, 0.55, 0.41}
\definecolor{lime-green}{rgb}{0.2, 0.8, 0.2}
\definecolor{teal-blue}{rgb}{0.21, 0.46, 0.53}

\def\dbFP#1{%
  {\color{salmon}\rule{\fpeval{#1/\inspectionSizePerOurDetectorNum*\barwidth} cm}{\barheight}}}
\def\dbTP#1{%
  {\color{lime-green}\rule{\fpeval{#1/\inspectionSizePerOurDetectorNum*\barwidth} cm}{\barheight}}}
\def\dbNW#1{%
  {\color{teal-blue}\rule{\fpeval{#1/\inspectionSizePerOurDetectorNum*\barwidth} cm}{\barheight}}}

\usepackage{multirow}
\usepackage{colortbl}
\definecolor{light-gray}{gray}{0.9}

\usepackage{tcolorbox}
\newenvironment{answerbox}{
\begin{tcolorbox}[colback=blue!5!white,colframe=blue!5!white,arc=0mm,left=1.5mm,right=1.5mm,top=0mm,bottom=0mm]
}
{
\end{tcolorbox}
}

\usepackage[usestackEOL]{stackengine}

\usepackage{listings}
\usepackage{xcolor}
\usepackage{colortbl}

\definecolor{celadon}{rgb}{0.67, 0.88, 0.69}
\definecolor{codegreen}{rgb}{0,0.6,0}
\definecolor{codegray}{rgb}{0.5,0.5,0.5}
\definecolor{codepurple}{rgb}{0.58,0,0.82}
\definecolor{backcolour}{rgb}{0.95,0.95,0.92}

\lstdefinestyle{mystyle}{
  backgroundcolor=\color{backcolour},
  commentstyle=\color{codegreen},
  keywordstyle=\color{magenta},
  numberstyle=\tiny\color{codegray},
  stringstyle=\color{codepurple},
  basicstyle=\fontsize{7pt}{8pt}\ttfamily, %
  breakatwhitespace=false,
  breaklines=true,
  captionpos=b,
  keepspaces=true,
  numbers=left,
  numbersep=5pt,
  showspaces=false,
  showstringspaces=false,
  showtabs=false,
  tabsize=2
}
\definecolor{pastelGreen}{rgb}{0.72, 0.90, 0.64}

\usepackage{xcolor}
\definecolor{celadon}{rgb}{0.67, 0.88, 0.69}
\def\code#1{\texttt{#1}}

\theoremstyle{definition}

\definecolor{backcolour_ql}{rgb}{0.91, 0.98, 0.99}
\lstdefinelanguage{codeql}{
    keywordstyle=\color{blue},
    backgroundcolor=\color{backcolour_ql},
    alsodigit = {-},
    keywords = {
      int, string, predicate, not,
      where, select, from, instanceof, and, this, or, import, class, extends},
    morestring=[b]",
    morecomment=[l]{//},
}

\usepackage{enumitem}
\setlist[itemize]{leftmargin=*}
\setlist[enumerate]{leftmargin=*}

\begin{document}

\title{Analyzing Quantum Programs with LintQ: A Static Analysis Framework for Qiskit}

\author{Matteo Paltenghi}
\orcid{0000-0003-2266-453X}
\affiliation{%
  \institution{University of Stuttgart}
  \city{Stuttgart}
  \country{Germany}
}
\email{mattepalte@live.it}

\author{Michael Pradel}
\orcid{0000-0003-1623-498X}
\affiliation{%
  \institution{University of Stuttgart}
  \city{Stuttgart}
  \country{Germany}
}
\email{michael@binaervarianz.de}

\renewcommand{\shortauthors}{Paltenghi and Pradel}

\begin{abstract}
As quantum computing is rising in popularity, the amount of quantum programs and the number of developers writing them are increasing rapidly.
Unfortunately, writing correct quantum programs is challenging due to various subtle rules developers need to be aware of.
Empirical studies show that 40--82\% of all bugs in quantum software are specific to the quantum domain.
Yet, existing static bug detection frameworks are mostly unaware of quantum-specific concepts, such as circuits, gates, and qubits, and hence miss many bugs.
This paper presents \approachName{}, a comprehensive static analysis framework for detecting bugs in quantum programs.
Our approach is enabled by a set of abstractions designed to reason about common concepts in quantum computing without referring to the details of the underlying quantum computing platform.
Built on top of these abstractions, \approachName{} offers an extensible set of \nPatternDetectors{} analyses that detect likely bugs, such as operating on corrupted quantum states, redundant measurements, and incorrect compositions of sub-circuits.
We apply the approach to a newly collected dataset of \nProgramsSelectedQiskit{} real-world Qiskit-based quantum programs, showing that \approachName{} effectively identifies various programming problems, with a precision of \allTPOverAllWarningsLintQBestCheckers{} in its default configuration with the \nCheckersAboveFiftyPrecision{} best performing analyses.
Comparing to a general-purpose linter and two existing quantum-aware techniques shows that almost all problems (\percTPsMissedByAllCompetitors{}) found by \approachName{} during our evaluation are missed by prior work.
\approachName{} hence takes an important step toward reliable software in the growing field of quantum computing.

\end{abstract}



\setcopyright{rightsretained}
\acmDOI{10.1145/3660802}
\acmYear{2024}
\copyrightyear{2024}
\acmSubmissionID{fse24main-p691-p}
\acmJournal{PACMSE}
\acmVolume{1}
\acmNumber{FSE}
\acmArticle{95}
\acmMonth{7}
\received{2023-09-28}
\received[accepted]{2024-04-16}

\maketitle

\makeatletter
\lst@AddToHook{PreSet}{\normallineskiplimit=0pt}
\makeatother

\section{Introduction}

Given the rising interest in quantum computing, ensuring the correctness of quantum software is increasingly important.
Studies on bugs in quantum computing platforms~\cite{paltenghiBugsQuantumComputing2022} and quantum programs~\cite{luoComprehensiveStudyBug2022a} show that many bugs in such software are problems specific to the quantum computing domain.
For example, \citet{paltenghiBugsQuantumComputing2022} and \citet{luoComprehensiveStudyBug2022a} report that 40\% and 82\%, respectively, of the bugs found in quantum software are due to quantum-specific bug patterns.
Detecting bugs in quantum programs is especially important because many bugs silently lead to unexpected results, which may be hard to spot due to the probabilistic results of quantum computations.

Unfortunately, popular bug detection tools, such as CodeQL~\cite{avgustinovQLObjectorientedQueries2016}, Pylint~\cite{PylintCodeAnalysis}, Flake8~\cite{Flake8YourTool}, Infer~\cite{calcagno2015moving}, and ErrorProne~\cite{ErrorProne}, are unaware of quantum computing.
These tools consist of two parts:
First, a framework that provides a set of abstractions to reason about general properties of programs, such as data flow and control flow.
Second, a set of analyses built on top of the framework, each of which detects a particular kind of bug.
This design has proven effective for general-purpose bug detection, as it allows to reuse the framework for different analyses.
Yet, the abstractions provided by these frameworks are not sufficient to reason about quantum-specific concepts, such as quantum gates, quantum circuits, and quantum registers.

\begin{figure}[t]
  \begin{minipage}[b]{.93\textwidth}
  \begin{lstlisting}[language=Python,escapechar=|]
# Create a quantum registers and a classical register
qreg = QuantumRegister(4)
creg = ClassicalRegister(3)
# Create a quantum circuit
circ = QuantumCircuit(qreg, creg)|\label{line:motiv_create_circuit}| # Bug 1: Oversized circuit
# Add gates and measurements to the circuit
for i in range(3):
    circ.h(i)
circ.measure(qreg[0], creg[0]) |\label{line:motiv_measure}|
circ.ry(0.9, qreg[0])|\label{line:motiv_op_after_measure}| # Bug 2: Operation after measurement
circ.measure([0, 1, 2], creg)
# Execute the circuit on a simulator
backend_sim = Aer.get_backend("qasm_simulator")
results = backend_sim.run(transpile(circ, backend_sim), shots=1024).result()
\end{lstlisting}
\begin{picture}(0,0)
  \put(262,30)
  {\includegraphics[height=3cm]{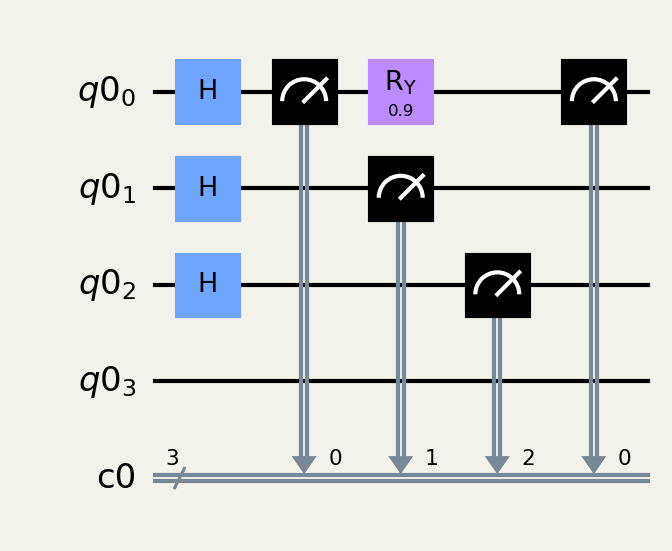}}
\end{picture}
\vspace{-1em}
\end{minipage}
\caption{Example of a quantum program with two bugs.}
\label{fig:bug_example}
\end{figure}

As a motivating example, Figure~\ref{fig:bug_example} shows a buggy quantum program.
The program is based on Qiskit, a popular quantum computing platform, where a quantum program is Python code that uses a specific library.
The code creates a circuit with a quantum register and a classical register, and then applies a sequence of gates and measurements to the circuit.
The image on the right of the figure shows a graphical representation of the circuit.
Despite being a simple program, the code contains two bugs.
First, the code creates a circuit with four qubits, but then uses only three of them.
Oversizing a circuit is strongly discouraged in quantum computing, because it wastes resources and because current hardware offers only a limited amount of qubits.
Second, the code measures the state of qubit~0 at line~\ref{line:motiv_measure}, and afterwards applies a gate to the same qubit at line~\ref{line:motiv_op_after_measure}.
Unfortunately, due to the properties of quantum mechanics, the measurement destroys the quantum state, and hence, the program feeds a collapsed state to the \code{ry} operation.
Finding such bugs in an automated bug detector requires the ability to reason about quantum-specific concepts.

Applying a general-purpose static bug detector, e.g., Pylint~\cite{PylintCodeAnalysis}, to the code in Figure~\ref{fig:bug_example} does not reveal the quantum-specific bugs.
Recently, first techniques aimed at quantum programs have been proposed.
One of them, QSmell~\cite{qihongchenSmellyEightEmpirical2023} relies on dynamic analysis for most of its checks, and hence, is inherently harder to apply to real-world programs than a static analysis.
Another one, QChecker~\cite{zhaoQCheckerDetectingBugs2023}, operates directly on the AST representation of quantum programs, but does not provide a general framework that abstracts over the details of the underlying quantum computing platform.
Besides their conceptual limitations, neither QSmell\footnote{Here and also in our empirical evaluation, we refer to the static subset of QSmell's checks because getting arbitrary quantum programs to execute is non-trivial, e.g., due to unresolved dependencies and user input expected by a program.} nor QChecker detects the bugs in Figure~\ref{fig:bug_example}, showing that there is a need for a comprehensive static bug detection framework for quantum programs.

This paper introduces \approachName{}, a static analysis framework for detecting bugs in quantum programs.
The approach is enabled by two key contributions.
First, \approachName{} offers a set of abstractions of common concepts in quantum computing, such as circuits, gates, and qubits.
These abstractions lift code written based on a specific quantum computing API, such as Qiskit, onto a higher level of abstraction.
Second, we implement on top these abstractions an extensible set of \nPatternDetectors{} analyses, each of which queries the code for a particular kind of quantum programming problem.
To benefit from prior work on general-purpose program analysis, the approach builds on an existing analysis framework, CodeQL~\cite{avgustinovQLObjectorientedQueries2016}.

Applying \approachName{} to the code in Figure~\ref{fig:bug_example} leads to warnings about the two bugs.
To reach this conclusion, the framework first represents the different elements of the program using our quantum-specific abstractions.
For example, this representation expresses the fact that the circuit created at line~\ref{line:motiv_create_circuit} has four qubits.
An analysis checking for oversized circuits then uses this information to determine that the circuit only uses the first three of the four qubits.
To find the second bug, an analysis aimed at warning about operations to a qubit applied after measuring the qubit builds upon the fact that our framework indicates that lines~\ref{line:motiv_measure} and~\ref{line:motiv_op_after_measure} operate on the same qubit.
Importantly, none of the analyses need to reason about specific API calls in the Python code, but instead reasons about the quantum program at the level of \approachName{}'s abstractions, which greatly simplifies the implementation of analyses.

We evaluate our approach by applying it to a novel dataset of \nProgramsSelectedQiskit{} real-world, Qiskit-based quantum programs.
The analyses built on top of our framework identify various problems in these programs.
Manually inspecting a sample of \nInspectedWarnings{} warnings from \nPatternDetectors{} analyses shows that \approachName{} identifies \nTruePositivesLintQ{} legitimate programming problems.
Moreover, when using the default and recommended configuration of \approachName{} with \nCheckersAboveFiftyPrecision{} analyses, it achieves a precision of \allTPOverAllWarningsLintQBestCheckers{} (\nTruePositivesLintQBestCheckers{} true positives out of \nInspectedWarningsBestCheckers{} warnings).
We reported \TPReported{} problems, \nConfirmed{} of which have already been confirmed or even fixed.
Our evaluation also shows that implementing an analysis on top of our abstractions is relatively simple, with an average of only \avgLoCQuery{} LoC per analysis, and that the analysis time is reasonable, with an average of \avgPerProgramEvaluationTimeSec{} seconds per program.

In summary, this work makes the following contributions:
\begin{itemize}
  \item A comprehensive framework for quantum program analysis, which provides reusable abstractions to reason about quantum software.
  \item \NPatternDetectors{} analyses implemented on top of these abstractions, which focus on programming problems reported in existing work on quantum-specific bugs~\cite{zhaoIdentifyingBugPatterns2021,kaulUniformRepresentationClassical2023}, or mentioned in GitHub issues and on StackExchange~\cite{egretta.thulaAnswerWhyDoes2023, OptimizeSwapBeforeMeasurePassDrops}.
  \item A novel dataset of \nProgramsSelectedQiskit{} real-world, Qiskit-based quantum programs, which is the largest such dataset and hence, may serve as a basis for future work.
  \item A thorough evaluation of the effectiveness of the analyses, showing that the approach in its default configuration finds real-world issues with a precision of \allTPOverAllWarningsLintQBestCheckers{} while taking only \avgPerProgramEvaluationTimeSec{} seconds to analyze a program.

\end{itemize}

\section{Background}
\label{sec:background}

\subsection{Quantum Programming}

Several quantum programming languages have been proposed, such as Qiskit~\cite{qiskit}, Cirq~\cite{developersCirq2021}, Q\#~\cite{svoreEnablingScalableQuantum2018}, and Silq~\cite{bichselSilqHighlevelQuantum2020}.
Quantum programs, also called quantum circuits, are expressed as a sequence of operations, called quantum gates, applied to individual qubits.
In Figure~\ref{fig:bug_example}, we show a Qiskit program with four qubits and one classical bit, represented as horizontal lines, whereas the gates are shown as boxes or colored vertical lines.
A special type of gate, called measurement gate, is used to measure the state of a qubit and store the result in a classical bit.
The measurement gate, represented in black in the figure, produces a certain bit, either 0 or 1, with probabilities determined by the qubit state.
Once the circuit has been defined, it is sent to a backend that executes it, typically a simulator or a real quantum computer.
Note that a measurement has the important side-effect of destroying the quantum state.
To sufficiently characterize its output, the circuit is executed multiple times, called \textit{shots}, each time measuring the qubit.

To define and run quantum programs, developers rely on a quantum computing platform.
A popular approach is to implement a platform on top of Python, as done by Qiskit~\cite{qiskit}, Cirq~\cite{developersCirq2021}, Tket~\cite{sivarajahKetRetargetableCompiler2020}, and Pennylane~\cite{bergholmPennyLaneAutomaticDifferentiation2020}, so that a quantum program is essentially Python code that uses a specific library.
Most platforms, with some noteworthy exceptions~\cite{bichselSilqHighlevelQuantum2020, paradisUnqompSynthesizingUncomputation2021, weigoldEncodingPatternsQuantum2021} describe programs on the level of circuits, and \approachName{} focuses on the analysis of quantum circuits.
In particular, \approachName{} focus on code written using the Qiskit quantum computing platform due to its popularity in both practice~\cite{darganTopQuantumProgramming2022a} and research~\cite{paltenghiBugsQuantumComputing2022, zhaoBugs4QBenchmarkReal2021, luoComprehensiveStudyBug2022a, fortunatoQMutPyMutationTesting2022, fortunatoMutationTestingQuantum2022, wangQDiffDifferentialTesting2021,paltenghiMorphQMetamorphicTesting2023}.

\subsection{Static Analysis with CodeQL}

\begin{figure}[t]

\end{figure}

\begin{figure}[t]
  \begin{minipage}[b]{.93\textwidth}
  \begin{minipage}[b]{.47\textwidth}
    \begin{lstlisting}[
      language=codeql, numbers=left, escapechar=&]
import python &\label{line:import_python}&
from If ifstmt, Stmt pass
where pass = ifstmt.getStmt(0) and &\label{line:predicate}&
  pass instanceof Pass
select ifstmt, "This 'if' statement is redundant."\end{lstlisting}
    \caption{Example CodeQL query to find code redundant if statements.}
    \label{fig:example_codeql}
  \end{minipage}
  \hfill
  \begin{minipage}[b]{.45\textwidth}
    \includegraphics[width=\textwidth]{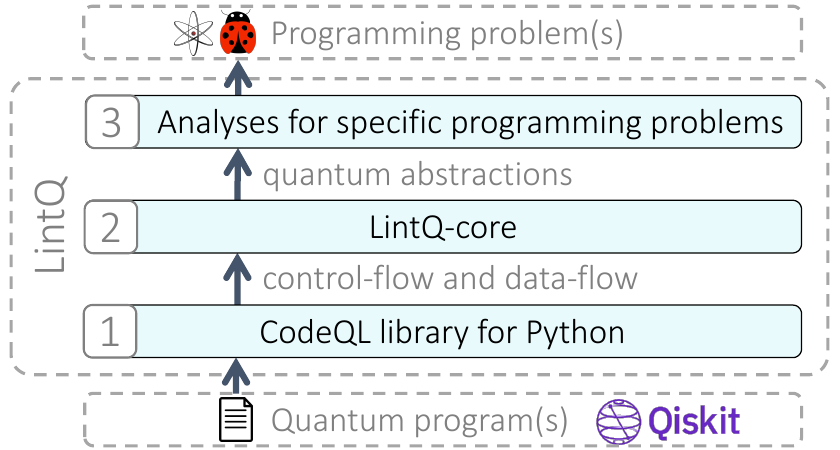}
    \caption{Overview of \approachName{}.}
    \label{fig:method_overview}
  \end{minipage}
  \end{minipage}
\end{figure}

CodeQL~\cite{avgustinovQLObjectorientedQueries2016}\footnote{\url{https://codeql.github.com/}} is a popular engine for static analysis.
It extracts facts from a program, such as its syntactic structure, data flow, and control flow, and then stores them in a relational database, which can be queried with the QL logic language.
A query refers to classes, which we call \emph{abstractions}, and their \emph{predicates}, which represent relations between the abstractions.
As a simple example, Figure~\ref{fig:example_codeql} shows a CodeQL query that finds redundant if statements in Python.
In the \code{from} section of the query, we define which program elements we consider, namely all if statements (\emph{If}) and all statements (\emph{Stmt}). Then, the \code{where} section restricts the query and focuses on those if statements that contain the \code{pass} keyword in their body. Finally, the \code{select} section specifies the warning and its message.
At line~\ref{line:predicate}, we see how the \emph{If} abstraction offers a \emph{getStmt} predicate, which connects the if statement with its first statement in the ``if true'' block.
Although Python is the language of choice of many quantum computing platforms~\cite{QiskitQiskit2021, developersCirq2021, sivarajahKetRetargetableCompiler2020}, the built-in QL library for Python extracts no information regarding quantum computing concepts, such as the quantum registers, the qubit position in a register, the quantum gates, or the difference between a state manipulation and a measurement, further motivating the need for \approachName{}.

\section{Overview of \approachName{}}
\label{sec:overview}

Figure~\ref{fig:method_overview} shows an overview of \approachName{}, which takes a quantum program written in Qiskit as its input, and then outputs warnings about quantum-specific programming problems in this program.
\approachName{} is a static analysis framework realized in three stages:
\begin{enumerate}
  \item The existing static analysis engine CodeQL (Section~\ref{sec:background}) extracts general information about Python code, such as control flow paths, data flow facts, and how to resolve imports.
  \item The core of our approach, called \core{}, represents the behavior of the quantum program using a set of reusable quantum programming abstractions, such as qubits, gates, and circuits.
  Section~\ref{sec:quantum_abstractions} describes these abstractions in detail.
  The key benefit provided by this stage is to lift the program representation from a large and diverse set of Python constructs and Qiskit APIs into a smaller set of reusable abstractions.
  \item An extensible set of analyses builds on the abstractions to identify programming problems.
  Each analysis is formulated as a query over facts provided by CodeQL and \core{}, which allows for writing concise yet precise analyses.
  Section~\ref{sec:target_bug_patterns} describes \nPatternDetectors{} analyses in detail.
\end{enumerate}

Stages~2 and~3 of \approachName{} are the main technical contributions of our work.
As Qiskit is a Python library, building on top of CodeQL in Stage~1 allows us to reuse its abilities at reasoning about Python programs.
At the same time, CodeQL does not have any knowledge of quantum programming, which is why we introduce \core{} in Stage~2.

\section{Quantum Abstractions}
\label{sec:quantum_abstractions}

\core{} provides a set of abstractions that represent concepts commonly found in quantum programs.
The motivation for introducing these abstractions is that quantum programming platforms, such as the Qiskit, typically offer a wide range of APIs to express quantum computations.
An alternative to \core{} would be to define analyses directly w.r.t.\ these APIs, which would require each analysis to consider the diversity of the Python language and the Qiskit APIs.
As an example to illustrate this diversity, consider how a program may refer to qubits.
First, there are multiple Python constructs for this purpose, including a single integer literal, e.g., \code{1}, a single integer variable, e.g., \code{qubit\_idx}, a
sequence of integer literals or integer variables, e.g., \code{[0, 1, 2]}, and expressions that retrieve a value from a variable that holds a qubit register, e.g., \code{qreg[2]}.
Second, all the above references to qubits may occur in various code locations.
For example, Qiskit offers over 50 functions to add different kinds of gates to a circuit.
Each of these functions expects references to qubits at one or more argument positions.
Instead of considering the full diversity of Python and Qiskit in each analysis, \core{} lifts quantum programs into more general abstractions.
These abstractions enable us to write concise analyses that reason about quantum computing concepts instead of a low-level API that implements these concepts.

\begin{figure}
  \centering
  \includegraphics[width=0.95\textwidth]{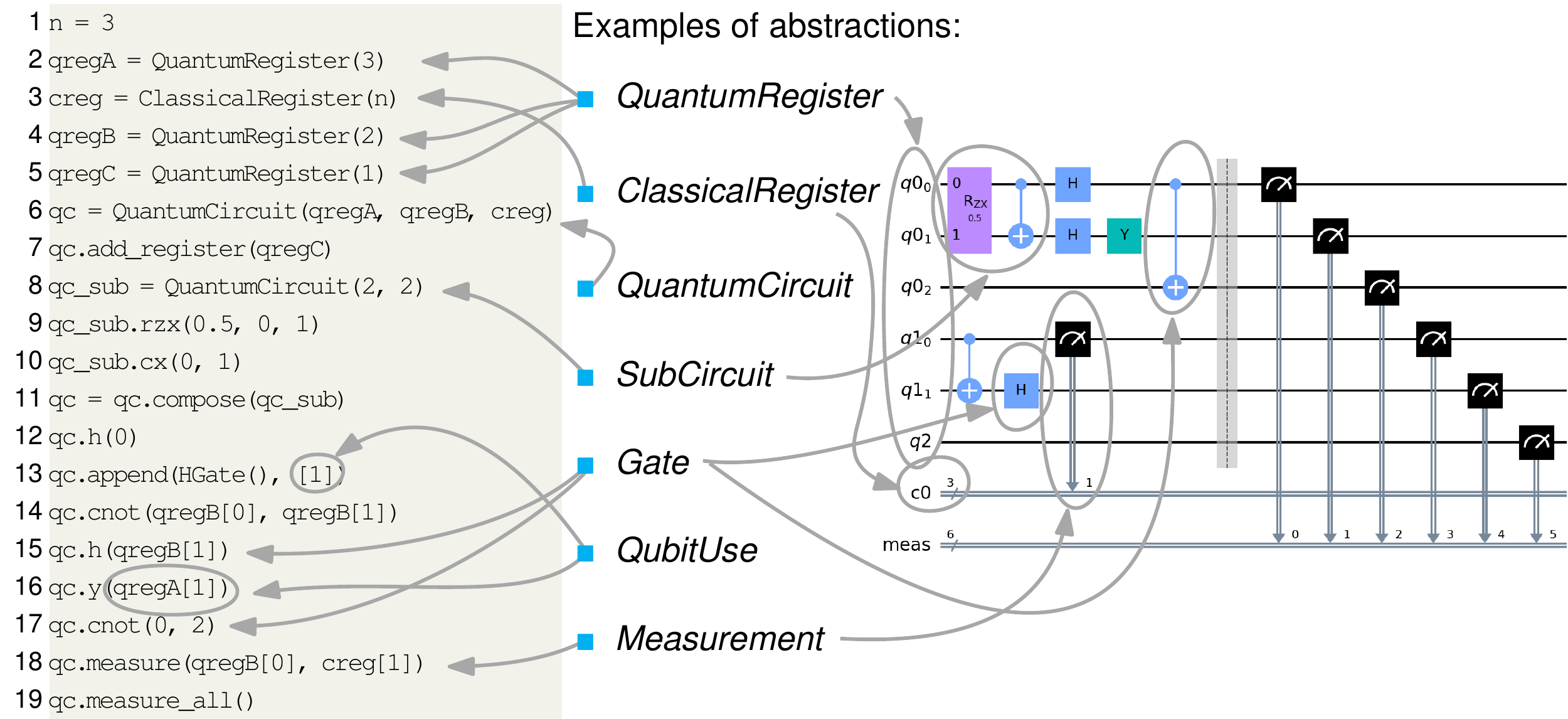}
  \caption{Examples of \approachName{}'s abstractions and how they are represented in the program and the circuit.}
  \label{fig:abstraction_example}
\end{figure}

Figure~\ref{fig:abstraction_example} gives an overview of the abstractions offered by \core{} and illustrates through an example how they relate to the Qiskit API.
Each abstraction corresponds to a CodeQL class that offers a number of predicates.
These predicates express relationships between abstractions, e.g., to reason about the quantum register that stores the qubit a gate operates on, or properties of the specific abstraction, e.g. the size of a register.
Although some of our abstraction borrow names used in the quantum circuit model~\cite{deutschQuantumComputationalNetworks1997, nielsenQuantumComputationQuantum2002} and in Qiskit, the abstractions represent more general concepts than simple types or API calls.
For example, as illustrated above, a single qubit or gate can be referred to in various ways, but they all share the same abstractions.
To clearly distinguish between the abstractions and the Qiskit API, we show abstractions in italics, e.g., \emph{QuantumCircuit} and Python/Qiskit code in monospace, e.g., \code{QuantumCircuit(qreg, creg)}.

\subsection{Registers}

The central storage facility of quantum programs are registers.
Lines~2 to~7 of Figure~\ref{fig:abstraction_example} illustrate different ways of creating registers and associating them with circuits.
For example, registers can be created explicitly by calling the respective constructors (lines~2 to~5), but also implicitly by passing the size of a register to a newly created circuit (line~8).
Reasoning about the size of a register requires us to identify the integer constant that determines the register size, as illustrated in line~1.
\approachName{} models registers and their relationships with the remaining program using two abstractions: \emph{QuantumRegister} and \emph{ClassicalRegister} for storing qubits and bits, respectively.
The register abstraction is defined by a data flow node corresponding to the register's allocation site in the source code, e.g., a call to a constructor \code{QuantumRegister}.
The abstraction offers predicates to retrieve the size of the register and the circuit it is associated with.
For example, the \emph{getSize} predicate of the \emph{ClassicalRegister} created at line~3 returns \code{3} because \approachName{} tracks the value of the variable \code{n} (line~1) and its relationship with the register.

\subsection{Quantum Circuits}

A quantum circuit describes a sequence of instructions that operate on data stored in registers.
Qiskit provides several APIs for creating circuits, composing larger circuits from smaller ones, and for associating circuits with other program elements, e.g., registers and gates.
For example, lines~6 and~8 of Figure~\ref{fig:abstraction_example} create two circuits and then add one as a sub-circuit into the other.

\approachName{} models circuits using the \textit{QuantumCircuit} abstraction.
The predicates offered by this abstraction include \emph{isSubcircuitOf}, which allows for checking whether a circuit is a sub-circuit of another one.
The \emph{getNumberOfQubits} and \emph{getNumberOfClassicalBits} predicates track the size of each circuit, even when it uses multiple registers, such as in line~6 or when the registers are added later in the code, such as in line~7.
Finally, \emph{getAGate} and \emph{getAQuantumRegister} express the relationship between a circuit and any of its gates or quantum registers.

The \textit{QuantumCircuit} abstraction is lifted from different source-code constructs: (1) calls to the \code{QuantumCircuit} constructor, (2) any call to a user-defined function returning a quantum circuit, (3) any built-in constructor of parametrized circuits, such as \code{EfficientSU2}, (4) any unknown object created via an external function call that uses methods specific to quantum circuits, such as \code{to\_gate}, \code{to\_instruction}, \code{assign\_parameters}, (5) any copy of an existent circuit created by calling \code{copy} on another circuit, (6) a call to \code{transpile}, which returns a new version if the circuit that is compatible with the instruction set and connectivity of the target quantum computer.
All of these cases are modeled as subclasses of the \textit{QuantumCircuit} abstraction, giving analyses the option to refer to specific kinds of circuits.
For example, \textit{TranspiledCircuit} abstraction can be used to enforce rules about circuits once they have been transpiled.

\subsection{Subcircuits and Composition}
To create larger circuits from smaller ones, quantum programming platforms offer APIs to compose circuits.
\approachName{} models how different circuits are composed with each other via the \emph{isSubcircuitOf} predicate.
The predicate \emph{isSubCircuitOf} tracks via dataflow analysis all those quantum circuits that flow into the \code{append} and \code{compose} methods of a quantum circuit object, and it identifies the object on which the method is called as the \emph{parent} circuit and the circuit passed as the argument as its \emph{subcircuit}.
In addition, \core{} models two other cases where a circuit is not yet explicitly composed into another one, but is likely to be a subcircuit: (a) when the circuit is returned by a function and (b) when an entire circuit is converted into an atomic instruction or gate via the \code{to\_instruction} or \code{to\_gate} methods.

\subsection{Quantum Operators: Reversible and Irreversible}

Instructions in quantum programs are expressed via quantum operators being added to a circuit.
Intuitively, a quantum operator is any function that manipulates the quantum state by acting on the values stored in one or more qubits.
There are two main types of quantum operators: reversible and irreversible.
Because irreversible operators, such as measurements, destroy the quantum state, they are typically placed at the end of a quantum program.
\core{} represents quantum operators using the \emph{QuantumOperator} abstraction, which is a superclass of the \emph{Gate}, \emph{Measurement}, and \emph{Reset} abstractions.

The \emph{Gate} abstraction represents reversible quantum operators, such as the Hadamard gates used in lines~12 and~13 of Figure~\ref{fig:abstraction_example}.
Overall, there are several dozens of different APIs for creating gates and many more for connecting gates with other parts of a quantum program, e.g., the qubits a gate operates on.
To enable analyses to reason about gates without repeatedly listing all gate-related APIs, \core{} offers the \emph{Gate} abstraction, which captures all gates and their properties.
The abstraction provides predicates to reason about a gate's relations to other program elements.
For example, the \emph{getQuantumCircuit} predicate relates a gate to the circuit it is added to, and the \emph{getATargetQubit} predicate allows for reasoning about the qubit a gate operates on.
For illustration, consider the control-not gate created at line~14 of Figure~\ref{fig:abstraction_example}.
The \emph{getATargetQubit} predicate returns the fact that this gate operates on the qubits stored at indices 0 and 1 of the quantum register created at line~4.

To represent irreversible quantum operators, \core{} offers multiple abstractions: \emph{Measurement} and \emph{MeasurementAll} to represent measurements of a single qubit and all qubits in a register, respectively; \emph{Reset} for operations that reset a qubit to the $|0\rangle$ state; and \emph{Initialize} for operations that initialize one or more qubits with a vector of complex numbers.
In Figure~\ref{fig:abstraction_example}, \approachName{} creates measurement abstractions for the code at lines~18 and~19.

\subsection{Uses of Qubits and Classical Bits}

Quantum information stored in qubits typically is used and manipulated by multiple quantum operators.
At the level of the Qiskit API, uses of qubits come in various forms.
For example, as illustrated above, a program may refer to a qubit via an integer passed as an argument to a gate operation or via an index into an array that represents a register.
Reasoning about qubit uses is compounded by the fact that different gate operations use different parameter indices to refer to the qubits they operate on.
For example, when adding a controlled unitary gate to a circuit via \code{qc.cu(1, 2, 3,  4, 5, 6)}, then only the last two arguments refer to qubits, whereas the others are parameters of the gate.

To help analyses in precisely reasoning about qubit uses, \approachName{} offers the \emph{BitUse} abstraction, split in its quantum and classical subclasses: \emph{QubitUse} and \emph{ClbitUse}.
The \emph{QubitUse} abstraction uniquely identifies a used qubit based on the register a qubit is stored in and based on the integer index of a qubit in this register.
The \emph{BitUse} abstraction offers predicates to connect to other abstractions, e.g., for obtaining the gate where the (qu)bit is used, the register where the (qu)bit is stored, and the corresponding circuit.
The \emph{getAnIndex} and \emph{getAnAbsoluteIndex} predicates return the position of the qubit in the register and the position of the qubit in the circuit, respectively.
The latter predicate keeps track of all the registers added to the circuit before the current accessed register and shifts the index accordingly.
For our running example in Figure~\ref{fig:abstraction_example}, the \emph{QubitUse} abstractions represents each of the many references to qubits, such as the use qubit~1 of \code{qregA} at line~13 or the use of qubit~0 of register \code{qregB} at line~14.

\subsection{Quantum Data Flow}
\label{sec:quantum data flow}

An important property of quantum computations, which does not have a direct correspondence in classical computing, is the order in which quantum operators are applied to qubits.
However, the order of adding two quantum operators to a circuit does not necessarily imply that the operator added first is executed after another operator added to the circuit later.
Instead, the order of applying operations depends on the qubits the operators act on.
\core{} derives the ordering of two operators if and only if they act in the same qubit.
To this end, the approach uses the \emph{QubitUse} abstraction described above to check if the qubits that two operators $op_1$ and $op_2$ manipulate are the same, and if so, derives an ordering relation based on the order in which $op_1$ and $op_2$ are added into the circuit.
Following this reasoning for all quantum operators yields a partial order between quantum operators, that we call \emph{quantum data flow}, since it describes how data stored in qubits flows between quantum operators.
\approachName{} exposes this partial order to analyses via the \emph{mayFollow} predicate that relates two quantum operators if and only if the two are part of the same circuit and the first may follow the second according to derived quantum data flow.

For example, consider the control-not gate (line~14) and the measurement (line~18) in Figure~\ref{fig:abstraction_example}.
Since both operate on the same qubit belonging to the \code{qregB} register, \approachName{} exposes their ordering in the \emph{mayFollow} predicate.
In contrast, the analysis does not claim any order between the gates created at lines~15 and~16, as they act on different registers.

In addition to the \emph{mayFollow} predicate, \core{} offers: (i) \emph{mayFollowDirectly}, a variant where the two quantum operators are applied on the same qubit directly after one another, without any other operation in between, (i) \emph{sortedInOrder}, which checks whether three quantum operators may appear in the given order according to the quantum data flow.
Note that both are defined only for unambiguous quantum operator additions, i.e., when the approach knows for sure that the operators act on the same qubit.

\subsection{Access to Low-Level Constructs}

Although our abstractions help in writing concise analysis, the analysis developer is not limited to those, but can refer to lower-level constructs when necessary.
For example, an analysis could restrict a \emph{Gate} to refer to specific kind of gate or reason on any other specific parameter of an API, if needed.
Ultimately, the main benefit of \approachName{} is to avoid repeating the same low-level details across different analyses by capturing commonly required abstractions.

\subsection{Soundness and Precision: Unknown Quantum Operators}

Following the philosophy of existing linters and many other static analyzers~\cite{Livshits2015}, \approachName{} aims neither at full soundness nor at full precision.
Instead, the approach offers a pragmatic compromise between the two, with the goal of finding as many programming problems as possible without overwhelming developers with spurious warnings.

To prevent analyses from raising incorrect warnings due to general limitations of static analysis, \core{} explicitly model unknown information, which gives an analysis the option to (now) draw conclusions based on such information.
Specifically, \core{} exposes a \emph{QubitUse} only if the approach can unambiguously resolve both the register and the index.
For example, if a program applies a Hadamard gate with \code{qc.h(idx)} where \code{idx} is a variable obtained from user input, then \approachName{} leaves the qubit access unresolved so that analyses do not draw inaccurate conclusions.
To the same end, \core{} exposes a \emph{UnknownQuantumOperator} abstraction when the analysis framework cannot resolve some of the qubits used in that operator, e.g., \code{qc.cx(0, i)}, where the value of \code{i} is not statically known.
\approachName{} also considers functions that may extend a circuit as an \emph{UnknownQuantumOperator}.
We identify those as either: (i) a call to an unknown function, where a \emph{QuantumCircuit} flows in as an argument, (ii) a call to a function that directly modifies the \emph{QuantumCircuit} by referring to it via a global variable.

As a preprocessing step to reduce the number of \emph{UnknownQuantumOperators}, when reasoning about programs with loops, the framework unrolls loops that have a statically known number of iterations.
Such loops are relatively frequent in quantum programs, e.g., when the programmer applies the same gates multiple  times and specifies the loop bound with \code{range(1, 3)} or \code{range(4)}.
We limit unrolling to loops with at most ten iterations as a tradeoff between better modeling of the program and the risk of introducing too many new program elements, thus affecting the scalability of the approach.

\section{Analyses for Finding Quantum Programming Problems}
\label{sec:target_bug_patterns}

\begin{table}
  \caption{Analyses for finding quantum programming problems.}
  \newcommand{\vmyrow}{\vspace*{1mm}}
  \label{tab:checkers}
  \small
  \setlength{\tabcolsep}{2pt}
  \begin{tabular}{@{}lp{33em}r@{}}
    \toprule
    \vmyrow{}Analysis name & Description & Origin \\
    \midrule
    \rowcolor{light-gray}  \multicolumn{3}{l}{\vmyrow{} Measurement-related and gate-related problems:}\\
     DoubleMeas & Two measurements measure the same qubit state one after the other. & \cite{zhaoIdentifyingBugPatterns2021} \\
    OpAfterMeas & A gate operates on a qubit after it has been measured. & \cite{zhaoIdentifyingBugPatterns2021} \\
    MeasAllAbuse & Measurement results are stored in an implicitly created new register, even though another classical register already exists. & \cite{zhaoIdentifyingBugPatterns2021} \\
    CondWoMeas & Conditional gate without measurement of the associated register. & \cite{kaulUniformRepresentationClassical2023} \\
    \vmyrow{}ConstClasBit & A qubit is measured but has not been transformed. & \cite{kaulUniformRepresentationClassical2023} \\
    \rowcolor{light-gray} \multicolumn{3}{l}{\vmyrow{} Resource allocation problems:}\\
    InsuffClasReg & Classical bits do not suffice to measure all qubits. & \cite{zhaoIdentifyingBugPatterns2021} \\
    \vmyrow{}OversizedCircuit & The quantum register contains unused qubits. & \cite{user19571QuestionRemoveInactive2022} \\
    \rowcolor{light-gray} \multicolumn{3}{l}{\vmyrow{} Implicit API constraints:}\\
    GhostCompose & Composing two circuits without using the resulting composed circuit. & \cite{egretta.thulaAnswerWhyDoes2023} \\
    OpAfterOpt & A gate is added after transpilation. & \cite{OptimizeSwapBeforeMeasurePassDrops} \\
    OldIdenGate & Using a now-removed API to create an identity gate. & \cite{zhaoQCheckerDetectingBugs2023} \\
    \bottomrule
  \end{tabular}
\end{table}

To illustrate the usefulness of \core{}, the following present an extensible set of \nPatternDetectors{} analyses built on top of the abstractions provided by the framework.

\subsection{Methodology: Collecting a Catalogue of Bug Patterns}
To identify a set of quantum programming problems that can be detected by static analysis, we search through existing literature and developer discussions.
We collect literature that studies programming issues in quantum programs by querying both the ACM Digital Library and IEEE Xplore, looking for any work that contains both keywords ``quantum'' and ``bug'' in its metadata.
We apply a cutoff on the search results and inspect the top 50 results of each database, sorted by relevance, which yields 100 candidate papers.
Next, we exclude papers that match one of the following:
they are duplicate papers (\miniSurveyExclusionReasonDuplicate{});
they focus on hardware faults (\miniSurveyExclusionReasonHardwareRelated{});
they discuss only quantum computing concepts (\miniSurveyExclusionReasonOnlyMentioningQuantum{}) or give an overview of the field (\miniSurveyExclusionReasonFieldOverview{});
they are theory papers (\miniSurveyExclusionReasonTheoryPaper{});
they only use quantum computing methods and are not focused on bug detection (\miniSurveyExclusionReasonOnlyUsingQuantum{});
they focus quantum-related software, such as quantum computing platforms, but not on quantum programs (\miniSurveyExclusionReasonNotOnPrograms{});
they do not provide a list of bugs or issues (\miniSurveyExclusionReasonNotBugList{});
or they require a specification of each quantum program, which typically is not available to a linter (\miniSurveyExclusionReasonSpecRequired{}).
We also exclude the dataset paper of Bugs4Q~\cite{zhaoBugs4QBenchmarkExisting2023}, because we use it as a benchmark in our evaluation.
After this filtering, we are left with \miniSurveyIncludedWorks{} papers that contain a list of programming issues or bug patterns in quantum programs.
To complement the list of bug patterns from the literature, we collect \miniSurveyPatternsFromDevDiscussion{} additional patterns that we identified in developer discussions on StackOverflow and in GitHub issues using an approach similar to prior work~\cite{luoComprehensiveStudyBug2022a}.

Inspecting the selected papers and discussions, we find a total of \miniSurveyTotalBugPatternsNoDuplicates{} unique bug patterns for which we have clear examples of the problem to be identified.
We exclude patterns for the following reasons:
implementing an accurate analysis requires knowledge of the exact hardware the program will run on (\miniSurveyBugPatternRequireHardwareKnowledge{});
they require a specification for each quantum program, e.g., describing the ``correct'' gate, which is generally not available to a linter (\miniSurveyBugPatternSpecRequired{});
they require runtime information (\miniSurveyBugPatternRuntimeInfo{});
they require abstractions not available in LintQ (\miniSurveyBugPatternLintqModelingLimitations{});
the described pattern is not an issue anymore in the current Qiskit release (\miniSurveyBugPatternNotAnIssue{});
or they look for rare combinations of APIs that hardly appear in our large evaluation dataset (\miniSurveyBugPatternTooApiSpecific{}).
After this filtering, we obtain a list of \miniSurveyTotalAnalysesImplemented{} analyses to implement in \approachName{}, which are listed in Table~\ref{tab:checkers}.
Note that, although the ultimate goal of LintQ is to find bugs, we acknowledge that some of the patterns could also be considered code smells or anti-patterns.
The following describes each analysis in detail by introducing the problem and by then describing how to query for instances of the problem using the \approachName{} abstractions.

\subsection{Measurement-Related and Gate-Related Problems}

\textbf{Double measurement.}
Any two subsequent measurements on the same qubit produce the same classical result, making the second measurement not only redundant but also a possible sign of unintended behavior or a misunderstanding of the properties of quantum information.
Figure~\ref{fig:bug_and_query_redundant_measurement} (left) shows an example of the problem.
\textit{Analysis}: The query to spot this problem is shown in Figure~\ref{fig:bug_and_query_redundant_measurement} (right).
It searches for two consecutive measurements of the same qubit by checking whether the two operations are directly adjacent w.r.t.\ the order derived from quantum data flow (Section~\ref{sec:quantum data flow}).
Note that simply relying on the integer to spot two gates operating on the same qubit is ineffective since they might refer to the same position but in two different registers.
To avoid this problem, the analysis relies on the \textit{mayFollowDirectly} predicate provided by \core{}, which leads to simple and concise analysis.

\begin{figure}[t]
  \begin{minipage}[b]{.93\textwidth}
  \begin{minipage}[b]{.44\textwidth}
  \begin{lstlisting}[language=Python, numbers=left]
circuit = QuantumCircuit(3, 3)
circuit.ccx(0, 1, 2)
circuit.measure(0, 0)
circuit.measure(2, 2)
# Problem: Qubit 0 already measured
circuit.measure(0, 1)
\end{lstlisting}
  \end{minipage}
  \hfill
  \begin{minipage}[b]{.52\textwidth}
  \begin{lstlisting}[
language=codeql, numbers=left, escapechar=&]
from Measurement m1, Measurement m2, int q
where mayFollowDirectly(m1, m2, q)
select m2, "Redundant measurement on the same qubit"\end{lstlisting}
\end{minipage}
\end{minipage}
\caption{Redundant measurement example (left) and its analysis (right).}
\label{fig:bug_and_query_redundant_measurement}
\end{figure}

\textbf{Operation after measurement.} When a qubit is measured, its quantum state collapses to a classical value, either 0 or 1. %
The measurement operation thus effectively destroys the quantum state.
Any subsequent operation after the measurement acts on a destroyed quantum state, which is unlikely to be the intended behavior.
Figure~\ref{fig:bug_and_query_op_after_measurement} (left) shows an example of the problem, where qubit 0 is measured and then a Pauli-Z gate is applied on it.
Note how this case differs from a redundant measurement, where the clash is between two measurements of the same qubit, whereas here it is between a measurement and a gate.
\textit{Analysis}: The query searches for a quantum gate that is applied on a qubit, which has just been measured.
Note that a trivial check of the usage of API calls that syntactically happen one after the other is insufficient, e.g., because the control flow may be more complex and the two calls could refer to two different qubits.
Instead, our query ensures that the two operations are applied both to the same qubit belonging to the same register thanks to the \code{mayFollowDirectly} predicate, as shown in Figure~\ref{fig:bug_and_query_op_after_measurement} (right).
The query excludes cases where applying the gate depends on a classical bit that resulted from a measurement, since this is a common pattern in quantum programs and the measurement preceding the gate could be required.

\begin{figure}[t]
  \begin{minipage}[b]{.93\textwidth}
  \begin{minipage}[b]{.5\textwidth}
  \begin{lstlisting}[language=Python, numbers=left]
qc = QuantumCircuit(2, 2)
qc.h(1)
qc.cx(1, 0)
qc.measure(0, 0)
qc.measure(1, 1)
qc.z(0) # Problem: Qubit 0 has collapsed
qc.measure(0, 0) \end{lstlisting}
  \end{minipage}
  \hfill
  \begin{minipage}[b]{.43\textwidth}
  \begin{lstlisting}[
    language=codeql, numbers=left, escapechar=&]
from Measurement m, Gate g, int q
where
  mayFollowDirectly(m, g, q)
  and not g.isConditional()
select gate, "Gate after measurement on qubit " + q \end{lstlisting}
\end{minipage}
\end{minipage}
  \caption{Operation after measurement example (left) and its analysis (right).}
  \label{fig:bug_and_query_op_after_measurement}
\end{figure}

\textbf{Measure all abuse.} In Qiskit, the API call \code{measure\_all} with default arguments is used to measure all the qubits of a program and store the result in a classical register that is generated on the fly.
Calling \code{measure\_all} on a circuit that already contains a classical register may cause a silent problem since the output string would include additional output registers, while the original classical register would likely end up being empty, initialized with all zeros.
Figure~\ref{fig:bug_and_query_measure_all} (left) shows an example, where the output bitstring has four bits instead of two, because of the newly added register.
Note that although the developer might still be able to correctly interpret the longer string, it is a waste of register space and might lead to unexpected results.
\textit{Analysis}: The query searches for a use of \code{measure\_all} on a \code{QuantumCircuit} that has a classical register.
This chain of relationships is handled by \approachName{} abstractions, as shown in Figure~\ref{fig:bug_and_query_measure_all} (right).
Note that, thanks to our abstractions, \approachName{} knows that the circuit has classical bits even if a \code{ClassicalRegister} object is not explicitly instantiated, but it gets this information indirectly when creating the circuit, i.e., \code{QuantumCircuit(2, 2)}.
Ultimately, we use an auxiliary predicate to ensure that the \code{measure\_all} gate indeed creates a new register, which happens when it gets called with default arguments,
i.e. whenever the argument is not \code{add\_bits=False}.

\textbf{Conditional gate without measurement.} In quantum programming, conditional gates play a crucial role in introducing conditional behavior into quantum circuits.
A conditional gate is applied to a target qubit only when a condition expressed through a classical bit is satisfied.
For example, \code{qc.h(0).c\_if(creg, 0)} applies the \code{h} gate only if the classical register \code{creg} contains the value \code{0}.
However, applying a conditional gate without any preceding measurement that stores a value into the classical bit(s) used in the condition essentially means a constant condition, which usually is not the programmer's intention.
\textit{Analysis}: The query searches for a conditional gate in a circuit in which all measurements are applied after the conditional gate, i.e., no preceding measurement exists.
For better precision, the query excludes warning raised on circuits involved in circuit compositions, because those circuits might be composed with others that have a preceding measurement.

\textbf{Constant classical bit.}
Whenever a qubit is manipulated, it is impossible to know its value without measuring it.
In contrast, if a qubit is never modified, its state remains in the initial default state, i.e., 0, and thus any measurements of it will certainly return a constant value.
\textit{Analysis:} The query searches for a measurement on a qubit for which there is no preceding gate applied on it.
To reduce false positives, it also excludes cases where the circuit has \emph{Subcircuits} or an \emph{UnknownQuantumOperator}.

\begin{figure}[t]
  \begin{minipage}[b]{.93\textwidth}
  \begin{minipage}[b]{.48\textwidth}
  \begin{lstlisting}[language=Python, numbers=left]
qc = QuantumCircuit(2, 2)
qc.h(q[0])
qc.cx(q[0], q[1])
# Problem: Implicitly creates a new classical register
qc.measure_all()
job = execute(qc,backend,shots=1000)
result = job.result().get_counts(qc)
# output: {'00 00': 487, '11 00': 513}\end{lstlisting}
  \end{minipage}
  \hfill
  \begin{minipage}[b]{.47\textwidth}
    \begin{lstlisting}[
      language=codeql, numbers=left, escapechar=&]
from
  QuantumCircuit c, MeasurementAll m
where c = m.getQuantumCircuit() and
  c.getNumberOfClassicalBits() > 0
  and m.createsNewRegister()
select m, "measure_all() with classical register"\end{lstlisting}
  \end{minipage}
\end{minipage}
  \caption{Measure all abuse example (left) and its analysis (right).}
  \label{fig:bug_and_query_measure_all}
\end{figure}

\subsection{Resource Allocation Problems}

The current generation of quantum computers are still limited in terms of qubits and gates, thus the use of resources must be carefully managed to avoid wasting them.
At the same time, enough resources must be allocated for the quantum state to evolve and being measured correctly.

\textbf{Insufficient classical register.} This problem happens when we define a quantum program that uses more qubits than those that can be measured in the classical register allocated in the beginning.
For example, the problem arises when the developer allocates a classical register with only two bits and then works on three qubits, i.e., \code{QuantumCircuit(3, 2)}.
\textit{Analysis}: The query searches for circuits with the number of qubits greater than the number of classical bits.
The \textit{QuantumCircuit} abstraction is used to reason about the number of classical and quantum bits, but also to check with the predicate \textit{isSubCircuit} that the circuit is not used as a sub-circuit.
This is necessary to reduce false positives, since sub-circuits are often used legitimately without a classical register.

\textbf{Oversized circuit.} This problem happens when a program allocates a quantum register that is larger than the number of qubits actually used.
Given the high cost of implementing a single qubit in hardware, when this issue happens it implies a waste of resources.
The motivating example in Figure~\ref{fig:bug_example} shows an instance of this problem, where the program allocates a quantum register of size four but uses only three of the qubits.
To the best of our knowledge, this work is the first to describe this programming problem.
\textit{Analysis:} As shown in Figure~\ref{fig:bug_and_query_oversized_quantum_circuit} (bottom), the query scans all gates used in a quantum circuit and raises a warning if any of the slots in the quantum register is not used (line~\ref{line:not_for_each_qubit}).
To ensure precision, the query checks several situations where no warning should be reported, e.g., circuits with an unknown register size or an unknown gate, and circuits that have sub-circuits.

\begin{figure}[t]
  \centering
  \begin{minipage}{0.93\textwidth}
  \begin{lstlisting}[
    language=codeql, numbers=left, escapechar=\&]
from QuantumCircuitConstructor circ, int numQubits
where
  // the circuit has a number of qubits
  numQubits = circ.getNumberOfQubits() and numQubits > 0 and
  // there is one qubit position not accessed by any gate
  not exists(QubitUse bu, int i | i in [0 .. numQubits - 1] | &\label{line:not_for_each_qubit}&
    bu.getAnAbsoluteIndex() = i and
    bu.getAGate().getQuantumCircuit() = circ
  ) and
  // the circuit has no (unknown) sub-circuits
  not exists(SubCircuit sub | sub.getAParentCircuit() = circ) and &\label{line:not_usage_as_subcrcuit}&
  // there is no initialize op, because it can potentially touch all qubits
  not exists(Initialize init | init.getQuantumCircuit() = circ) and
  // all its registers have well-known size
  not exists(QuantumRegisterV2 reg | reg = circ.getAQuantumRegister() and not reg.hasKnownSize()) and &\label{line:not_unknown_reg_size}&
  // there are no unknown quantum operators
  not exists(UnknownQuantumOperator unkOp | unkOp.getQuantumCircuit() = circ) &\label{line:not_unknown_quantum_op}&
select circ, "Circuit has unused qubits"\end{lstlisting}
\end{minipage}
\caption{Analysis to detect oversized circuits (see Figure~\ref{fig:bug_example} for an example).}
  \label{fig:bug_and_query_oversized_quantum_circuit}
\end{figure}

\subsection{Implicit API Constraints}

The Qiskit API imposes several implicit constraints that quantum program should respect.

\textbf{Ghost composition.} When two circuits are composed with a \code{compose} call, the final merged circuit is the return value of the method call.
That is, failing to use this return value leads to what we call a ``ghost'' composition, where the sub-circuit is not added at all.
The problem is difficult to detect since it does not lead to any crash, but likely leads to unintended output.
\textit{Analysis}: The query searches for a call of \code{compose} on a quantum circuit, where the return value is not assigned to a variable or used otherwise.
The query does not raise a warning in case the \code{compose} call has a \code{inplace=True} parameter, which would make the composition correct.

\textbf{Operation after transpilation.}
Transpilation is a procedure to adapt the quantum program to the specific architecture of the quantum computer, and it involves optimization procedures.
One particularly insidious optimization pass, which is active with optimization level~3, is ``Optimize Swap Before Measure'', because it removes any swap gate at the end of a circuit if the qubits it acts on are not measured.
Thus, if the transpilation is done before adding the final measurements, this leads to a silent bug with unexpected results, because the swap gates get removed.
\textit{Analysis:} The query searches for a gate added to a circuit that was transpiled with optimization level~3.
Thanks to the \textit{TranspiledCircuit} abstraction, the reasoning on all transpiled circuits can be done with a single abstraction, which also offers a predicate to check the optimization level.

\textbf{Old identity gate.}
The identity gate is a special gate that does not change the quantum state.
Since Qiskit v0.23.0, the identity gate API call \code{qc.iden(index)} has been superseded by two new APIs: \code{qc.i(index)} and \code{qc.id(index)}.
The use of deprecated APIs is a potential source of future problems, and it is thus important to detect it.
\textit{Analysis:} The query searches for the deprecated \code{iden} call in a quantum circuit. This analysis uses the existing \textit{QuantumCircuit} abstraction, and could be easily extended to cover other deprecated API calls.

\section{Evaluation}
\label{sec:evaluation}

We seek to answer the following research questions:
\begin{itemize}
  \item \textbf{RQ1:} How effective are the \core{} abstractions at supporting the analyses?
  \item \textbf{RQ2:} Which bugs are detected by \approachName{} in real-world quantum programs?
  \item \textbf{RQ3:} What precision and recall does \approachName{} offer, and what causes false positives?
  \item \textbf{RQ4:} How does \approachName{} compare to existing static analyses?
  \item \textbf{RQ5:} How efficient is \approachName{}?
\end{itemize}

\subsection{Dataset}
\label{sec:dataset}

The evaluation applies \approachName{} to a newly collected dataset of \nProgramsSelectedQiskit{} real-world quantum programs gathered from GitHub.
To gather Qiskit programs, which are the targets of \approachName{}, we query the GitHub search API for .py and .ipynb files that contain \code{import qiskit} or \code{from qiskit import} on \dataDatasetMiningQiskit{}.
Because some repositories contain generated and duplicate programs, we manually inspect the 30 repositories with most files returned, and remove repositories with generated or duplicate programs, coming mainly from companion repositories of research projects~\cite{wangQDiffDifferentialTesting2021,paltenghiMorphQMetamorphicTesting2023,paltenghiBugsQuantumComputing2022,zhaoBugs4QBenchmarkReal2021}.
This filter reduces the dataset from \KeepOnlyPyAndIpynb{} to \RemoveBlacklistedRepos{} programs.
Next, we convert the notebooks to Python scripts using \code{nbconvert}~\cite{NbconvertConvertNotebooks}.
Finally, we discard (i) files that cannot be parsed, (ii) duplicates based on the hash of the file content, (iii) files not including the search keywords in their code but, e.g., in a comment, (iv) code that is part of Qiskit itself. %
To the best of our knowledge, the resulting \nProgramsSelectedQiskit{} programs are the largest dataset of real-world quantum programs to date.
The resulting dataset includes \TotalLinesOriginalPrograms{} lines of code, excluding blank lines and comments.
Figure~\ref{fig:gate_calls_entire_dataset} shows the frequency of the top-10 most popular gates in the dataset.
The most popular gate are the Pauli-X gate (\code{\topFirstGateName{}}), followed by the Hadamard (\code{\topSecondGateName{}}), and the controlled-X gate (\code{\topThridGateName{}}).
On average, a program has \avgGateCallsPerFileNoMeasure{} gate calls and \avgMeasureCallsPerFile{} calls to the \code{measure} API.

\subsection{RQ1: Effectiveness of Abstractions}

The primary goal of \approachName{}'s abstraction is to allow for writing concise analyses.
To quantify to what extent we reach this goal, we compute the number of lines of code (LoC) needed to implement the various analyses.
With an average of only \avgLoCQuery{} LoC, varying from \minLoCQuery{} to \maxLoCQuery{} LoC, we confirm that our analyses are short.
Note that we limit each line to 80 characters and that we also include the select clause, which provides an informative error message and therefore is longer than shown in the example analyses in Section~\ref{sec:target_bug_patterns}.

\begin{figure}[t]
  \begin{minipage}[b]{.3\textwidth}
    \centering
    \includegraphics[width=\textwidth]{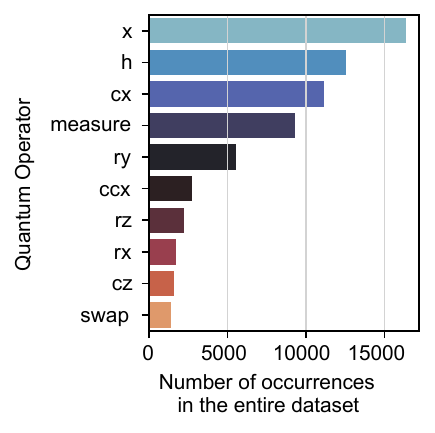}
    \caption{Frequency of most popular gates in our dataset.}
    \label{fig:gate_calls_entire_dataset}
  \end{minipage}
  \hfill
  \begin{minipage}[b]{.65\textwidth}
  \centering
  \includegraphics[width=\textwidth]{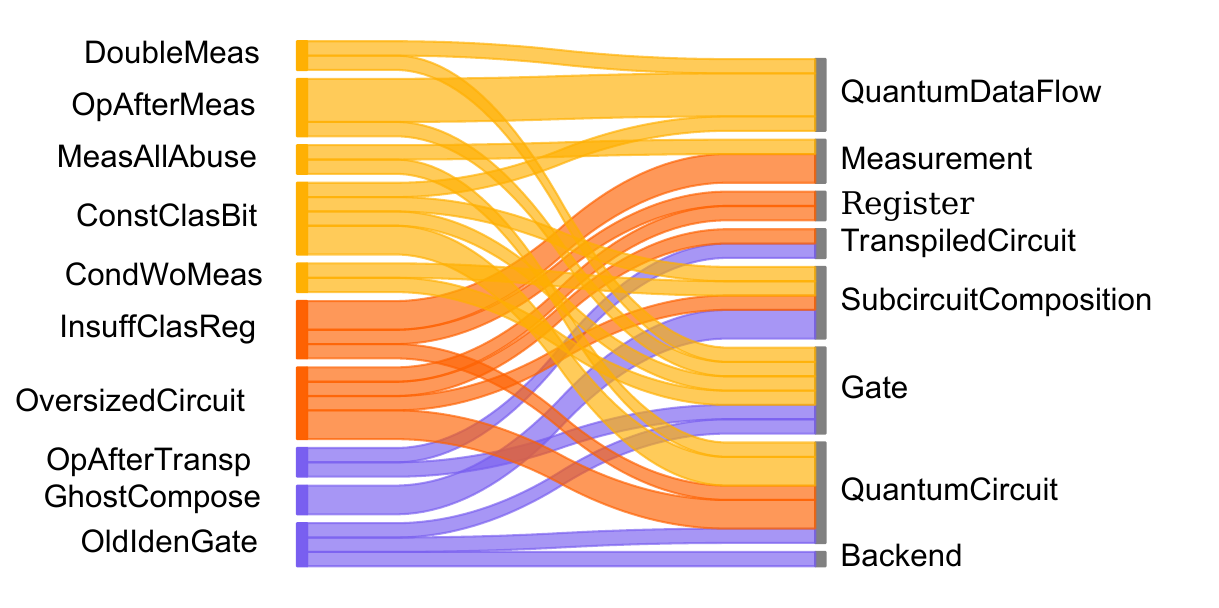}
  \caption{Mapping between analyses (left) and the abstraction families they use (right).}
  \label{fig:mapping_query_abstr}
  \end{minipage}
\end{figure}

To quantify the reusability of our abstractions, Figure~\ref{fig:mapping_query_abstr} illustrates what analyses (left) explicitly use what abstraction families (right).
The width of the connection between an analysis and a family of abstractions is proportional to the number of abstractions of that family that are used by the analysis.
For example, \emph{OpAfterMeas} uses the \emph{mayFollow} and \emph{sortedInOrder} predicates, which belong to the \emph{QuantumDataFlow} family of abstractions.
Each abstraction is used, on average, by \avgQueriesSupportedPerAbstraction{} queries, showing that the abstractions are effectively reused across multiple analyses, with the \emph{QuantumCircuit} being the most reused one.
Each analysis uses, on average, \avgAbstractionsUsedPerQuery{} abstractions, showing how few abstractions are needed to implement an analysis.
This is possible because the predicates often allow for indirectly obtaining information about other abstractions without explicitly declaring them in the query.
For example, the \emph{getNumberOfQubits} predicate of the \emph{QuantumCircuit} abstraction allows for indirect reasoning about the size of registers without using the \emph{QuantumRegister} abstraction.

\begin{answerbox}
  \textbf{Answer to RQ1}: \approachName{} abstractions are general enough to be reused across multiple analyses, which enables analyses with only \avgLoCQuery{} LoC, on average.
\end{answerbox}

\subsection{RQ2: Bugs Detected by \approachName{}}

Table~\ref{tab:precision_estimate} presents the number of warnings reported by the different analyses.
In addition to the absolute number of warnings, the third column also shows how many of all analyzed files have at least one warning produced by an analysis.
The results show that the number of warnings varies widely from analysis to analysis.
We evaluate the precision of analyses, i.e., their ability to pinpoint relevant problems, as opposed to false positives, in RQ3.

\begin{table*}
  \centering
  \small
  \caption{Examples of true positives. The ID refers to the supplementary material.}
  \setlength{\tabcolsep}{3pt}
  \begin{tabular}{@{}rllll@{}}
    \toprule
    ID & Analysis & Description & Manifestation & Status \\
    \midrule
    1 & DoubleMeas & Redundant measurement & Silent & Fixed \\
    2 & GhostCompose & Missing compose assignment in a test case & Silent & Fixed \\
    3 & GhostCompose & Missing compose assignment in a test case & Silent & Fixed \\
    4 & GhostCompose & Missing compose assignment in a test case & Silent & Fixed \\
    5 & MeasAllAbuse & Measurement creates overly long output & Silent & Fixed \\
    6 & MeasAllAbuse & Measurement creates overly long output & Silent & Confirmed \\
    7 & OpAfterMeas & Non idiomatic teleportation, lack of c\_if & Silent & Confirmed \\
    8 & InsuffClasReg & Unused qubit & Silent & Reported \\
    9 & OpAfterTransp & Measure gate added after transpilation & Silent & Reported \\
    10 & OpAfterTransp & Measure gate added after transpilation (Grover's algor.) & Silent & Reported \\
    11 & MeasAllAbuse & Measurement creates overly long output (Bell pair) & LongOut & Reported \\
    12 & MeasAllAbuse & Measurement creates overly long output (GHZ algor.) & LongOut & Reported \\
    13 & MeasAllAbuse & Measurement creates overly long output & LongOut & Reported \\
    14 & GhostCompose & Missing insertion of the inverse QFT (Shor's algor.) & Silent & Reported \\
    15 & MeasAllAbuse & Measurement creates overly long output (CTF competition) & Silent & Not Fix \\
    \bottomrule
    \end{tabular}

  \label{tab:bug_reports}
  \end{table*}

\textbf{Reporting Protocol.}
While inspecting a subset of the warnings that \approachName{} produces, we found \nTruePositivesLintQ{} bugs that should be fixed in the quantum programs.
Because adequately reporting bugs to developers is a time-consuming process, we report a subset of \TPReported{} of the true positives.
The subset is selected by prioritizing warnings based on their likelihood of eliciting a response from the developers and based on the relevance of the project.
Concretely, we consider the number of stars, contributors, issues, and the days since the last commit to identify projects where reporting issues is worthwhile.
We then review all the remaining true positives and after careful consideration, we identify some cases in which, although the programming issue is present, it is not worth the time of the developers to fix it.
The cases which we deemed not worth reporting issues are:
the repository is archived (\TPUnreportedBecauseArchived{}) or not maintained (\TPUnreportedBecauseUnmantained{}), e.g., with more than two years since the last commit;
the problem happens in a test case that runs despite the quantum circuit being buggy, because it tests a different feature (\TPUnreportedBecauseTestingOtherFeatures{});
the problem happens in a personal project, which was obviously not created with the intention of maintaining the code (\TPUnreportedBecauseIncompletePersonalProgramSketches{});
the problem happens in the interactive context of a notebook, where incomplete or buggy code is generally allowed for visualization purposes (\TPUnreportedBecauseIntermediateVisualization{}),
or in teaching material, which is purposefully buggy or incomplete (\TPUnreportedBecauseTeachingCode{}).

\textbf{Developer Responses.}
Table~\ref{tab:bug_reports} lists a selection of the reported bugs, along with a short description, the analysis involved, and how the problem manifests.
If the ``Manifestation'' column reports ``silent'', then the bug does not lead to any obvious sign of misbehavior, such as an exception, but instead either silently produces an incorrect result or causes unnecessary computation.
Finally, the last column shows which bugs we have already reported to the developers, and if available, their response.
In total, \nConfirmed{} of the bugs have already been confirmed or even fixed, and all but one of the remaining reports are still pending.

\textbf{Examples.}
We describe some examples of bugs found by \approachName{}.
Figure~\ref{fig:bug_example_redundant_measurement} shows code where a measurement is performed on a qubit that has just been measured.
The second measurement hence is redundant and causes unnecessary computation.
The developers confirmed and fixed this bug in response to our report.
As a second example, Figure~\ref{fig:bug_example_compose} shows some testing code where the states of two circuits are compared. However, the two circuits are identical  because the \code{compose} call is not properly assigned to the second circuit (line~\ref{line:compose_assignment}), and thus, no circuit contains the Toffoli gate (line~\ref{line:toffoli}), which should instead have been the subject of the test.
This bug was confirmed by the developers and fixed in response to our report.
Notably, the only ``not fix'' is because the repository belonged to a competition submission, and the developers preferred to keep the version that they submitted for historical reasons.
Interestingly, \approachName{} also finds a \code{measure\_all} bug in the official Qiskit tutorials (Id~1 in Table~\ref{tab:bug_reports}), which was confirmed by the developers, showing that even highly maintained quantum code may suffer from problems \approachName{} can find.

\begin{figure}
  \begin{minipage}[b]{.93\textwidth}
  \begin{subfigure}{0.45\textwidth}
    \centering
    \begin{lstlisting}[language=Python, firstnumber=1]
def inefficientNOT(inefficiencies: int, inp: str):
  qc = QuantumCircuit(1, 1)
  qc.reset(0)
  if inp == '1':
    qc.x(0)
  qc.barrier()
  for i in range(inefficiencies):
    print(i+1, "x gates added")
    qc.x(0)
  qc.barrier()
  qc.measure(0,0)
  trial = qc.measure(0,0)  # Bug\end{lstlisting}
  \vspace*{-.3em}
  \caption{Redundant measurement problem. {(ID: 2)}}
          \label{fig:bug_example_redundant_measurement}
  \end{subfigure}%
  \hfill
  \begin{subfigure}{0.5\textwidth}
    \centering
\begin{lstlisting}[language=Python, firstnumber=1, escapechar=|]
def test_linear_toffoli2(self):
  gate_x = np.array([[0, 1], [1, 0]])
  qc2 = QuantumCircuit(4)
  qc2.x(2)
  qc2.x(3)
  qc2.x(0)
  state1 = qclib.util.get_state(qc2)
  circ = QuantumCircuit(4)
  mc_gate(gate_x, circ, [3, 2, 1], 0)|\label{line:toffoli}|
  qc2.compose(circ, qc2.qubits) # Bug|\label{line:compose_assignment}|
  state2 = qclib.util.get_state(qc2)
  self.assertTrue(np.allclose(state1, state2))
\end{lstlisting}
    \vspace*{-.3em}
    \caption{Ghost composition problem. {(ID: 3)}}
    \label{fig:bug_example_compose}
  \end{subfigure}
  \vspace{.5em}
  \caption{Examples of bugs found by \approachName{}.}
  \label{fig:main}
  \vspace{.3em}
  \end{minipage}
\end{figure}

\begin{answerbox}
  \textbf{Answer to RQ2}: \approachName{} identifies problems in various real-world quantum programs, some of which were already confirmed and fixed by the respective developers.
\end{answerbox}

\subsection{RQ3: Precision and Recall}
\label{sec:precision recall}

\begin{table}
  \centering
  \caption{Warnings and precision of the \approachName{} analyses (left) and the result of manual inspection (right).}
  \small
  \label{tab:precision_estimate}
  \setlength{\tabcolsep}{9pt}
  \begin{tabular}{lp{1.75cm}p{0.9cm}|p{1cm}l@{}}
    \toprule
    Analysis Name & Tot. warnings & \% Files & Precision & FP / NW / TP \\
    \midrule
    DoubleMeas & 39 & 0.36\% & 72.0\% & \dbFP{1.60}\dbNW{1.20}\dbTP{7.20} 4/3/18 \\
    OpAfterMeas & 127 & 0.92\% & 100.0\% & \dbFP{0.00}\dbNW{0.00}\dbTP{10.00} 0/0/44 \\
    MeasAllAbuse & 22 & 0.26\% & 94.1\% & \dbFP{0.00}\dbNW{0.59}\dbTP{9.41} 0/1/16 \\
    ConstClasBit & 533 & 4.29\% & 48.3\% & \dbFP{3.50}\dbNW{1.67}\dbTP{4.83} 21/10/29 \\
    CondWoMeas & 46 & 0.22\% & 100.0\% & \dbFP{0.00}\dbNW{0.00}\dbTP{10.00} 0/0/28 \\
    InsuffClasReg & 3489 & 17.35\% & 34.8\% & \dbFP{3.33}\dbNW{3.18}\dbTP{3.48} 22/21/23 \\
    OversizedCircuit & 378 & 3.01\% & 50.0\% & \dbFP{2.76}\dbNW{2.24}\dbTP{5.00} 16/13/29 \\
    OpAfterTransp & 7 & 0.05\% & 100.0\% & \dbFP{0.00}\dbNW{0.00}\dbTP{10.00} 0/0/7 \\
    GhostCompose & 12 & 0.09\% & 66.7\% & \dbFP{0.00}\dbNW{3.33}\dbTP{6.67} 0/4/8 \\
    OldIdenGate & 46 & 0.37\% & 50.0\% & \dbFP{3.93}\dbNW{1.07}\dbTP{5.00} 11/3/14 \\
    \bottomrule
    \end{tabular}
\end{table}

\textbf{Precision}.
Because precision is crucial for practical adoption of static analyzers~\cite{flanaganExtendedStaticChecking2002, besseyFewBillionLines2010a, johnsonWhyDonSoftware2013}, we assess to what extent the analyses suffer from false positives.
We manually inspect a random sample of \inspectionSizePerOurDetector{} warnings for each analysis, or all produced warnings if that number is lower than \inspectionSizePerOurDetector{}.
Two of the authors, who are both experienced in static analysis and with quantum computing knowledge, independently inspect the warnings and then discuss them to reach a consensus.
After the initial inspection, a \percAgreement{} agreement was reached, and after the discussion, all disagreements were resolved.
Based on the agreement, we compile an annotation protocol and a single author proceeds to annotate more warnings up to reaching a statistically relevant sample with a confidence level of 90\% and a margin of error of 10\% for each of the \nPatternDetectors{} analyses, similar to related work~\cite{ghalebACheckerStaticallyDetecting2023}.

We categorize each warning into one of three categories.
A \textit{true positive} is a warning that reveals clearly incorrect behavior in the program.
Such incorrect behavior may result in a program crash, in incorrect output, or in an unnecessary performance degradation.
A \textit{noteworthy} warning is a potential problem where the analysis correctly detects an instance of the targeted programming problem, but we cannot certainly say whether the behavior is unexpected by the developer.
Finally, a \textit{false positive} is a warning reported despite the code being correct, which is typically caused by overly strong assumptions made by an analysis.
Based on this classification, we compute \emph{precision} as the percentage of true positives among all warnings.
That is, our notion of precision underestimates the true precision, as it includes noteworthy warnings in the denominator, but not in the numerator.

Table~\ref{tab:precision_estimate} illustrates the results of our manual inspection.
Each analysis identifies at least a few true positives.
The median precision across all analyses is \medianPrecisionLintQ{}.
The overall precision across all inspected warnings is \allTPOverAllWarningsLintQ{} (\nTruePositivesLintQ{} true positives out of \nInspectedWarnings{} inspected warnings).
In practice, we recommend to enable those analyses that produce sufficiently precise results for the usage environment, and to inspect warnings by high-precision analyses first.
For example, keeping only analyses with precision above 50\%, yields an overall precision of \allTPOverAllWarningsLintQBestCheckers{} from the remaining \nCheckersAboveFiftyPrecision{} analyses, which we recommend as the default configuration for \approachName{}.

\textbf{Root causes of false positives}. To better understand the reasons for false positives, we discuss representative cases in the following.
False positives of the \emph{InsuffClasReg} happen when: (ii) the circuit has more qubits than classical bits because of the presence of ancilla qubits, i.e., auxiliary storage used during a computation that does not need to be measured; (ii) the circuit is used as a submodule of a bigger circuit, thus it is not responsible of instantiating the classical bits.
Better distinguishing between ancilla qubits and missed classical bits remains as a challenge for future work.
The \emph{OversizedCircuit} analysis causes false positives when
the circuit has sub-circuits generated via a function call (e.g., \code{qc.append(QFT(3), qargs=[0, 1, 2])}), which \approachName{} currently does not track, thus making the circuit appear underused.

\textbf{Recall}.
Since we do not know the ground truth of all bugs in the \nProgramsSelectedQiskit{} real-world quantum programs, we cannot compute the recall of \approachName{} on this large dataset.
Instead, we use Bugs4Q~\cite{zhaoBugs4QBenchmarkReal2021, luoComprehensiveStudyBug2022a}\footnote{\url{https://github.com/Z-928/Bugs4Q-Framework}}, an existing benchmark of 42 quantum bugs.
We run \approachName{} on the 42 buggy files, manually inspect the warnings by using the same annotation procedure applied in the rest of this work, and then check how many of \approachName{}'s true positives match a known bug in Bugs4Q.
\approachName{} raises four true positive warnings (two by the \emph{OpAfterMeas}, one by \emph{MeasAllAbuse}, and one by \emph{OldIdenGate}) that correspond to three known bugs in Bugs4Q.
Thus, the recall of \approachName{} is \recallBugsFourQ{} (3/42).
While this number may seem low, it is actually higher than the recall of popular static bug detectors on Defect4J~\cite{justDefects4JDatabaseExisting2014}, which prior work has measured to be between 1\% and 3\%, depending on the bug detector~\cite{habibHowManyAll2018}.
Interestingly, \approachName{} also finds some problems in the benchmark code beyond the known bugs of the benchmark.
For example, the \emph{InsuffClasReg} analysis raises several true positive warnings because some circuits do not have any classical register, which is likely due to the fact that the examples are incomplete code snippets gathered from issues and forum questions.

\begin{answerbox}
  \textbf{Answer to RQ3}: Our analyses have an overall precision of \allTPOverAllWarningsLintQ{}, or when turning off low performing analyses, \allTPOverAllWarningsLintQBestCheckers{} from the remaining \nCheckersAboveFiftyPrecision{} analyses which we recommend as the default configuration for \approachName{}.
  The recall is \recallBugsFourQ{} on the Bugs4Q benchmark.
  Common root causes of false positives are ancilla qubits and insufficient modeling of complex circuit compositions.
\end{answerbox}

\subsection{RQ4: Comparison with Prior Work}

Due to the young field of quantum software engineering, there are only few static analyzers aimed at quantum programs: (i) QSmell~\cite{qihongchenSmellyEightEmpirical2023}, which detects smells in quantum programs, (ii) QChecker~\cite{zhaoQCheckerDetectingBugs2023}, an AST-based static analysis tool for quantum programs, (iii) QCPG~\cite{kaulUniformRepresentationClassical2023}, a toolkit that extends Code Property Graphs~\cite{yamaguchiModelingDiscoveringVulnerabilities2014} to analyze quantum code.
Moreover, since the quantum programs we analyze are written in Python, we also compare with Pylint~\cite{PylintCodeAnalysis}, a popular linter for Python designed for classical software.
For QSmell we focus on their two static analysis-based detectors; for QChecker we use their eight AST-based checkers; for Pylint, we run the tool its default configuration.
Unfortunately, we had to drop QCPG from the comparison because the tool is not available yet.\footnote{Although the authors plan to release the tool, the repository states the code is undergoing export checks, as confirmed by email.}

\textbf{Prior work applied to problems found by \approachName{}}.
We run all three competitors on the programs where \approachName{} detects one of the \nTruePositivesLintQ{} true positives that we have manually confirmed (Section~\ref{sec:precision recall}), and we check which of them the existing techniques detect.
QSmell, QChecker, and Pylint raise \nWarningsOnLintQTPwithQSmell{}, \nWarningsOnLintQTPwithQChecker{}, and \nWarningsOnLintQTPwithPylint{} warnings, respectively.
We inspect each warning that is at the same line as one of the \approachName{} warnings, which corresponds to \nWarningsOnLintQTPwithQSmellOverlap{}, \nWarningsOnLintQTPwithQCheckerOverlap{}, and \nWarningsOnLintQTPwithPylintOverlap{} warnings, and we assess whether they refer to the same problem or coincidentally flag the same line.
For the QChecker warnings, we found that seven problems are found by both QChecker and \approachName{}, two of QChecker's warnings coincidentally flag the same line, and the remaining \nWarningsOnLintQTPwithQCheckerMissed{} are missed by QChecker and found only by \approachName{}.
For the Pylint warnings, 15 Pylint warnings correspond to the same issues flagged by \emph{OldIdenGate} (14) and \emph{OversizedCircuit} (1), 32 warnings coincidentally flag the same line, including conventions and coding style (29), and uses of missing APIs and arguments (3), and the remaining \nWarningsOnLintQTPwithPylintMissed{} warnings are missed by Pylint and found only by \approachName{}.
Overall, prior work can find only \percTPsFoundByAtLeastOneCompetitor{} (\nTPsFoundByAtLeastOneCompetitor{}/\nTruePositivesLintQ{}) of the true positives found by \approachName{}, overlooking the remaining \percTPsMissedByAllCompetitors{} (\nTPsMissedByAllCompetitors{}/\nTruePositivesLintQ{}).

\textbf{\approachName{} applied to problems found by prior work}.
We also study the opposite direction: How many of the warnings raised by the competitors are also raised by \approachName{}?
To this end, we run each quantum-specific competitor and inspect a sample of up to ten warnings produced by each of their detectors.
For QSmell and QChecker we inspect, respectively, \nInspectedWarningsQSmellStatic{} warnings from \nDetectorsWithWarningsQSmellStatic{} detectors and \nInspectedWarningsQChecker{} warnings from \nDetectorsWithWarningsQChecker{} detectors.

For QSmell, we unfortunately found no true positives among the \nInspectedWarningsQSmellStatic{} inspected warnings.
The first detector (\emph{NC}) flags a file that has more \code{run} and \code{execute} calls, used to run a circuit on a simulator or real hardware, than \code{bind\_parameters} calls, used to convert any parametric gate into its concrete version before execution.
These warnings are false positives, for two reasons: (1) The warning is emitted also when there is a single \code{execute} call, which is normal for any circuit that uses only concrete gates. (2) The detector does not model the \code {assign\_parameters} API, which is a legitimate alternative to calling \code{bind\_parameters}.
The second detector (\emph{LPQ}) checks if there is a \code{transpile} API call without the argument \code{initial\_layout} set, since passing that argument is a good practice when running on a real quantum computer.
Again, all the inspected warnings are false positives, for two main reasons: (1) The missing \code{initial\_layout} argument is present with a simulator backend, which in practice has no hardware constraint to respect. (2) The rule considers any \code{transpile} calls, even those not belonging to Qiskit, which has no \code{initial\_layout} argument.

For QChecker, we found \nQcheckerTP{} true positives among the \nInspectedWarningsQChecker{} inspected warnings.
The true positives are raised by the \emph{Deprecated Order} detector, which flags a deprecated usage of the \code{iden}, analogously to our \emph{OldIdenGate} analysis.
However, QChecker also reports many false positives, primarly because it warns about any function call that includes the substring \code{iden}, which also happens in functions unrelated to the quantum library.
All \nQcheckerTP{} true positives found by QChecker are also detected by \approachName{}, because our \emph{OldIdenGate} analysis targets the same kind of problem.
The main difference is that \approachName{} raises fewer false positives, because it explicitly models gates, instead of relying on a text-based matching of the API name.

\begin{answerbox}
  \textbf{Answer to RQ4}: Overall, \percTPsMissedByAllCompetitors{} of the \nTruePositivesLintQ{} true positves issues found by \approachName{} are overlooked by the existing techniques, whereas the three true positives found by the existing quantum-specific techniques QSmell and QChecker are also found by \approachName{}.
  Pylint remains effective on mostly classical patterns, such as \emph{OldIdenGate}, but it misses all other quantum-specific problems found by \approachName{}.
\end{answerbox}

\subsection{RQ5: Efficiency}

We measure the time spent for analyzing all \nProgramsSelectedQiskit{} quantum programs with all \nPatternDetectors{} analyses.
All experiments are run on an Ubuntu machine with an Intel Xeon Silver 4214 CPU with 48 cores and 252 GB of RAM.
There are three main computational steps:
(i) Using CodeQL to build the database of facts about the Python code, which takes \totalDatasetCreationTimeMin{} minutes for all \nProgramsSelectedQiskit{} programs;
(ii) Compiling the query plan of the analyses, which takes \totalDatasetQueryCompilationTimeSec{} seconds; and
(iii) Running the analyses on the database, which takes \totalDatasetEvaluationTimeMin{} minutes.
Inspecting the computational cost of individual analyses shows that the two most expensive analyses are those that reason about the gate execution order, \emph{\topThreeDetectorRuleOne{}} and \emph{\topThreeDetectorRuleTwo{}}, which take \topThreeDetectorRuleOneTimeSec{} and \topThreeDetectorRuleTwoTimeSec{} seconds to evaluate, respectively.
Taking all three steps together, \approachName{} takes \avgPerProgramEvaluationTimeSec{} seconds per analyzed program.

\begin{answerbox}
  \textbf{Answer to RQ5}:
  With \avgPerProgramEvaluationTimeSec{} seconds per analyzed program, \approachName{} is sufficiently fast for a practical analysis.
\end{answerbox}

\section{Threats to Validity}

\paragraph{Internal Validity}
First, our analyses scan each program file individually, not considering the other files in the same repository.
Applying \approachName{} at the repository level may produce different warnings.
Second, we inspect only a subset of all warnings. To mitigate this thread, we sample the inspected warnings randomly and use a statistically relevant sample size.
Third, our implementation may contains bugs. To mitigate this risk, we implement test cases for the abstractions and the analyses, and we make our implementation publicly available as open-source.
Fourth, our literature review may have missed some relevant bug patterns. However, LintQ is designed to be extensible, i.e., additional bug patterns can be added to the framework in the future.
Fifth, the manual inspection of warnings is inherently subjective. To mitigate this, two authors participated in the process, collaboratively developing and agreeing on an annotation protocol. This protocol has been documented and is made available in our artifact.

\paragraph{External Validity}
First, while the abstractions of \approachName{} are designed to be also applicable to other quantum computing platforms, such as Cirq~\cite{developersCirq2021} and Tket~\cite{sivarajahKetRetargetableCompiler2020}, we cannot claim that our results generalize beyond Qiskit.
Second, we cannot guarantee that our results generalize to other quantum programs.
To mitigate this threat, we evaluate the approach on \nProgramsSelectedQiskit{} real-world programs, which represents the largest such dataset to date.

\section{Related Work}
\label{sec:related-work}

\paragraph{Quantum Software Testing}
\citet{miranskyyYourQuantumProgram2020} highlight quantum-specific debugging issues when working with quantum programs and discuss how classical solutions could be adapted to the quantum domain.
Regarding platform code, various approaches have been proposed, including differential testing~\cite{wangQDiffDifferentialTesting2021}, metamorphic testing~\cite{paltenghiMorphQMetamorphicTesting2023}, and fuzzing~\cite{xiaFuzz4AllUniversalFuzzing2024}.
However, they all focus on platform code and require executing the code, whereas \approachName{} focuses on application code and is based on static analysis.
Regarding application code, various techniques have been proposed.
QuanFuzz\cite{wangPosterFuzzTesting2021} tests a single algorithm with different inputs with the goal of maximizing branch coverage via a genetic algorithm.
Quito~\cite{aliAssessingEffectivenessInput2021a} relies on a program specification and statistical tests to evaluate the correctness of a single small program.
\citet{huangStatisticalAssertionsValidating2019} propose statistical approaches to evaluate assertions in a quantum program.
\citet{liProjectionbasedRuntimeAssertions2020} describe a projection-based runtime assertion scheme that allows for asserting in the middle of the circuit without affecting the tested state if the assertion is satisfied.
All these approaches assume to have single circuit programs that can be easily executed multiple times, which may not be the case in practice.
Moreover, they rely on executing the programs, whereas \approachName{} is based on static analysis.

\paragraph{Quantum Program Analysis}
Few analyses for quantum programs have been proposed so far, including QSmell~\cite{qihongchenSmellyEightEmpirical2023}, ScaffCC~\cite{javadiabhariScaffCCFrameworkCompilation2014}, QChecker~\cite{zhaoQCheckerDetectingBugs2023}, and QCPG~\cite{kaulUniformRepresentationClassical2023}.
QSmell mostly relies on dynamic analysis, and it focuses on code smells only.
ScaffCC is a compiler that performs a limited set of analyses using a new flavor of QASM, whereas we focus on Qiskit-based Python code.
QChecker is a static analysis tool that relies only on AST information, but does not provide a general framework to build new analyses and does not model any control flow.
QCPG~\cite{kaulUniformRepresentationClassical2023} extends Code Property Graphs~\cite{yamaguchiModelingDiscoveringVulnerabilities2014} to analyze quantum code in a single circuit, whereas \approachName{} is designed to analyze entire programs and models the composition of circuits, as well as unknown quantum operators.
There also exist quantum-specific program analyses, such as entanglement analysis~\cite{javadiabhariScaffCCFrameworkCompilation2014, perdrixQuantumEntanglementAnalysis2008} and automatic uncomputation~\cite{bichselSilqHighlevelQuantum2020, paradisUnqompSynthesizingUncomputation2021, xiaStaticEntanglementAnalysis2023}. However, most of them address a single problem each, whereas \approachName{} offers a set of general abstractions.%
Quantum abstract interpretation~\cite{yuQuantumAbstractInterpretation2021} and runtime assertions~\cite{liProjectionbasedRuntimeAssertions2020} are two techniques to assert properties of quantum computations.
They require manually crafted, algorithm-specific assertions, whereas \approachName{} does not require any prior knowledge of the program.
In summary, \approachName{} goes beyond the purely syntactic level and single-circuit approaches by providing reusable abstractions to build a wide range of analyses in realistic settings with multiple circuits and unknown quantum operators.

\paragraph{Datasets of Quantum Programs}
\citet{paltenghiBugsQuantumComputing2022} share a large dataset of bugs in quantum computing platforms. However, the focus of \approachName{} is on application code written in Qiskit, and not on platform code.
Two application-level datasets are QASMBench~\cite{liQASMBenchLowLevelQuantum2023}, which includes 48 programs written in OpenQASM, and work by \citet{longEquivalenceIdentityUnitarity2023}, which proposes a dataset of 63 Q\# programs.However, they are not suitable for our evaluation because they are not written in Python/Qiskit.
\citet{luoComprehensiveStudyBug2022a} and \citet{zhaoBugs4QBenchmarkReal2021} study the bugs in quantum computing programs in Qiskit, mainly collected from GitHub issues of the official Qiskit repository and StackOverflow questions.
In contrast, we present a much larger dataset of \nProgramsSelectedQiskit{} real-world programs, including many programs that are not part of the Qiskit repositories.

\paragraph{Domain-Specific API Modeling}
Previous work has modeled other specialized Python libraries, e.g., in machine learning~\cite{lagouvardosStaticAnalysisShape2020,bakerDetectFixVerify2022}, to spot bugs with static analysis.
Our work also relates to general static API misuse detectors~\cite{amannSystematicEvaluationStatic2019}, which mostly focuses on Java and traditional application domains.
Instead, we focus on the quantum domain, which comes with its own concepts and APIs to model.

\section{Conclusion}
We present \approachName{}, a framework for statically analyzing quantum programs and an extensible set of \nPatternDetectors{} analyses.
The approach introduces a set of abstractions that capture common concepts in quantum programs, such as circuits, gates, and qubits, as well as the relations between these concepts.
Thanks to these abstractions, analyses aimed at finding specific kinds of programming problems can be easily implemented in a few lines of code (\avgLoCQuery{} LoC).
To evaluate \approachName{}, we apply the approach to a novel dataset of \nProgramsSelectedQiskit{} quantum programs, and in its default configuration with \nCheckersAboveFiftyPrecision{} analyses, \approachName{} achieves a precision of \allTPOverAllWarningsLintQBestCheckers{} (\nTruePositivesLintQBestCheckers{} true positives out of \nInspectedWarningsBestCheckers{} warnings).

\section{Data Availability}

\approachName{}, our dataset, and all results are publicly available at \url{https://github.com/sola-st/LintQ} and archived at \url{https://zenodo.org/records/11095456}.

\section{Acknowledgments}

This work was supported by the European Research Council (ERC, grant agreement 851895), and by the German Research Foundation within the ConcSys, DeMoCo, and QPTest projects.

\bibliographystyle{ACM-Reference-Format}
\bibliography{phd-mattepalte,referencesMichael,referencesMore}


\begin{thebibliography}{56}


\ifx \showCODEN    \undefined \def \showCODEN     #1{\unskip}     \fi
\ifx \showDOI      \undefined \def \showDOI       #1{#1}\fi
\ifx \showISBNx    \undefined \def \showISBNx     #1{\unskip}     \fi
\ifx \showISBNxiii \undefined \def \showISBNxiii  #1{\unskip}     \fi
\ifx \showISSN     \undefined \def \showISSN      #1{\unskip}     \fi
\ifx \showLCCN     \undefined \def \showLCCN      #1{\unskip}     \fi
\ifx \shownote     \undefined \def \shownote      #1{#1}          \fi
\ifx \showarticletitle \undefined \def \showarticletitle #1{#1}   \fi
\ifx \showURL      \undefined \def \showURL       {\relax}        \fi
\providecommand\bibfield[2]{#2}
\providecommand\bibinfo[2]{#2}
\providecommand\natexlab[1]{#1}
\providecommand\showeprint[2][]{arXiv:#2}

\bibitem[Fla({[n.\,d.]})]%
        {Flake8YourTool}
 \bibinfo{year}{[n.\,d.]}\natexlab{}.
\newblock \bibinfo{title}{Flake8: {{Your Tool For Style Guide Enforcement}}
  \textemdash{} Flake8 6.0.0 Documentation}.
\newblock \bibinfo{howpublished}{https://flake8.pycqa.org/en/latest/}.
\newblock


\bibitem[Nbc({[n.\,d.]})]%
        {NbconvertConvertNotebooks}
 \bibinfo{year}{[n.\,d.]}\natexlab{}.
\newblock \bibinfo{title}{Nbconvert: {{Convert Notebooks}} to Other Formats
  \textemdash{} Nbconvert 7.2.9 Documentation}.
\newblock
  \bibinfo{howpublished}{https://nbconvert.readthedocs.io/en/latest/index.html}.
\newblock


\bibitem[Opt({[n.\,d.]})]%
        {OptimizeSwapBeforeMeasurePassDrops}
 \bibinfo{year}{[n.\,d.]}\natexlab{}.
\newblock \bibinfo{title}{{{OptimizeSwapBeforeMeasure}} Pass Drops {{Swap}}
  Gate (Even If There Is {{NO}} Measure after It) {$\cdot$} {{Issue}} \#7642
  {$\cdot$} {{Qiskit}}/Qiskit}.
\newblock \bibinfo{howpublished}{https://github.com/Qiskit/qiskit/issues/7642}.
\newblock


\bibitem[Pyl({[n.\,d.]})]%
        {PylintCodeAnalysis}
 \bibinfo{year}{[n.\,d.]}\natexlab{}.
\newblock \bibinfo{title}{Pylint - Code Analysis for {{Python}} |
  Www.Pylint.Org}.
\newblock \bibinfo{howpublished}{https://www.pylint.org/}.
\newblock


\bibitem[Qis(2021)]%
        {QiskitQiskit2021}
 \bibinfo{year}{2021}\natexlab{}.
\newblock \bibinfo{title}{Qiskit/Qiskit}.
\newblock \bibinfo{howpublished}{https://github.com/Qiskit/qiskit}.
\newblock


\bibitem[Ali et~al\mbox{.}(2021)]%
        {aliAssessingEffectivenessInput2021a}
\bibfield{author}{\bibinfo{person}{Shaukat Ali}, \bibinfo{person}{Paolo
  Arcaini}, \bibinfo{person}{Xinyi Wang}, {and} \bibinfo{person}{Tao Yue}.}
  \bibinfo{year}{2021}\natexlab{}.
\newblock \showarticletitle{Assessing the {{Effectiveness}} of {{Input}} and
  {{Output Coverage Criteria}} for {{Testing Quantum Programs}}}. In
  \bibinfo{booktitle}{\emph{2021 14th {{IEEE Conference}} on {{Software
  Testing}}, {{Verification}} and {{Validation}} ({{ICST}})}}.
  \bibinfo{pages}{13--23}.
\newblock
\showISSN{2159-4848}
\urldef\tempurl%
\url{https://doi.org/10.1109/ICST49551.2021.00014}
\showDOI{\tempurl}


\bibitem[Amann et~al\mbox{.}(2019)]%
        {amannSystematicEvaluationStatic2019}
\bibfield{author}{\bibinfo{person}{Sven Amann}, \bibinfo{person}{Hoan~Anh
  Nguyen}, \bibinfo{person}{Sarah Nadi}, \bibinfo{person}{Tien~N. Nguyen},
  {and} \bibinfo{person}{Mira Mezini}.} \bibinfo{year}{2019}\natexlab{}.
\newblock \showarticletitle{A {{Systematic Evaluation}} of {{Static API-Misuse
  Detectors}}}.
\newblock \bibinfo{journal}{\emph{IEEE Transactions on Software Engineering}}
  \bibinfo{volume}{45}, \bibinfo{number}{12} (\bibinfo{date}{Dec.}
  \bibinfo{year}{2019}), \bibinfo{pages}{1170--1188}.
\newblock
\showISSN{1939-3520}
\urldef\tempurl%
\url{https://doi.org/10.1109/TSE.2018.2827384}
\showDOI{\tempurl}


\bibitem[Avgustinov et~al\mbox{.}(2016)]%
        {avgustinovQLObjectorientedQueries2016}
\bibfield{author}{\bibinfo{person}{Pavel Avgustinov}, \bibinfo{person}{Oege de
  Moor}, \bibinfo{person}{Michael~Peyton Jones}, {and} \bibinfo{person}{Max
  Sch{\"a}fer}.} \bibinfo{year}{2016}\natexlab{}.
\newblock \showarticletitle{{{QL}}: {{Object-oriented Queries}} on {{Relational
  Data}}}. In \bibinfo{booktitle}{\emph{30th {{European Conference}} on
  {{Object-Oriented Programming}} ({{ECOOP}} 2016)}}
  \emph{(\bibinfo{series}{Leibniz {{International Proceedings}} in
  {{Informatics}} ({{LIPIcs}})}, Vol.~\bibinfo{volume}{56})},
  \bibfield{editor}{\bibinfo{person}{Shriram Krishnamurthi} {and}
  \bibinfo{person}{Benjamin~S. Lerner}} (Eds.). \bibinfo{publisher}{{Schloss
  Dagstuhl\textendash Leibniz-Zentrum fuer Informatik}},
  \bibinfo{address}{{Dagstuhl, Germany}}, \bibinfo{pages}{2:1--2:25}.
\newblock
\showISBNx{978-3-95977-014-9}
\showISSN{1868-8969}
\urldef\tempurl%
\url{https://doi.org/10.4230/LIPIcs.ECOOP.2016.2}
\showDOI{\tempurl}


\bibitem[Baker et~al\mbox{.}(2022)]%
        {bakerDetectFixVerify2022}
\bibfield{author}{\bibinfo{person}{Wilson Baker}, \bibinfo{person}{Michael
  O'Connor}, \bibinfo{person}{Seyed~Reza Shahamiri}, {and}
  \bibinfo{person}{Valerio Terragni}.} \bibinfo{year}{2022}\natexlab{}.
\newblock \showarticletitle{Detect, {{Fix}}, and {{Verify TensorFlow API
  Misuses}}}. In \bibinfo{booktitle}{\emph{2022 {{IEEE International
  Conference}} on {{Software Analysis}}, {{Evolution}} and {{Reengineering}}
  ({{SANER}})}}. \bibinfo{pages}{925--929}.
\newblock
\showISSN{1534-5351}
\urldef\tempurl%
\url{https://doi.org/10.1109/SANER53432.2022.00110}
\showDOI{\tempurl}


\bibitem[Bergholm et~al\mbox{.}(2020)]%
        {bergholmPennyLaneAutomaticDifferentiation2020}
\bibfield{author}{\bibinfo{person}{Ville Bergholm}, \bibinfo{person}{Josh
  Izaac}, \bibinfo{person}{Maria Schuld}, \bibinfo{person}{Christian Gogolin},
  \bibinfo{person}{M.~Sohaib Alam}, \bibinfo{person}{Shahnawaz Ahmed},
  \bibinfo{person}{Juan~Miguel Arrazola}, \bibinfo{person}{Carsten Blank},
  \bibinfo{person}{Alain Delgado}, \bibinfo{person}{Soran Jahangiri},
  \bibinfo{person}{Keri McKiernan}, \bibinfo{person}{Johannes~Jakob Meyer},
  \bibinfo{person}{Zeyue Niu}, \bibinfo{person}{Antal Sz{\'a}va}, {and}
  \bibinfo{person}{Nathan Killoran}.} \bibinfo{year}{2020}\natexlab{}.
\newblock \showarticletitle{{{PennyLane}}: {{Automatic}} Differentiation of
  Hybrid Quantum-Classical Computations}.
\newblock \bibinfo{journal}{\emph{arXiv:1811.04968 [physics,
  physics:quant-ph]}} (\bibinfo{date}{Feb.} \bibinfo{year}{2020}).
\newblock
\showeprint[arxiv]{1811.04968}~[physics, physics:quant-ph]


\bibitem[Bessey et~al\mbox{.}(2010)]%
        {besseyFewBillionLines2010a}
\bibfield{author}{\bibinfo{person}{Al Bessey}, \bibinfo{person}{Ken Block},
  \bibinfo{person}{Ben Chelf}, \bibinfo{person}{Andy Chou},
  \bibinfo{person}{Bryan Fulton}, \bibinfo{person}{Seth Hallem},
  \bibinfo{person}{Charles {Henri-Gros}}, \bibinfo{person}{Asya Kamsky},
  \bibinfo{person}{Scott McPeak}, {and} \bibinfo{person}{Dawson Engler}.}
  \bibinfo{year}{2010}\natexlab{}.
\newblock \showarticletitle{A Few Billion Lines of Code Later: Using Static
  Analysis to Find Bugs in the Real World}.
\newblock \bibinfo{journal}{\emph{Commun. ACM}} \bibinfo{volume}{53},
  \bibinfo{number}{2} (\bibinfo{date}{Feb.} \bibinfo{year}{2010}),
  \bibinfo{pages}{66--75}.
\newblock
\showISSN{0001-0782}
\urldef\tempurl%
\url{https://doi.org/10.1145/1646353.1646374}
\showDOI{\tempurl}


\bibitem[Bichsel et~al\mbox{.}(2020)]%
        {bichselSilqHighlevelQuantum2020}
\bibfield{author}{\bibinfo{person}{Benjamin Bichsel},
  \bibinfo{person}{Maximilian Baader}, \bibinfo{person}{Timon Gehr}, {and}
  \bibinfo{person}{Martin Vechev}.} \bibinfo{year}{2020}\natexlab{}.
\newblock \showarticletitle{Silq: A High-Level Quantum Language with Safe
  Uncomputation and Intuitive Semantics}. In
  \bibinfo{booktitle}{\emph{Proceedings of the 41st {{ACM SIGPLAN Conference}}
  on {{Programming Language Design}} and {{Implementation}}}}
  \emph{(\bibinfo{series}{{{PLDI}} 2020})}. \bibinfo{publisher}{{Association
  for Computing Machinery}}, \bibinfo{address}{{New York, NY, USA}},
  \bibinfo{pages}{286--300}.
\newblock
\showISBNx{978-1-4503-7613-6}
\urldef\tempurl%
\url{https://doi.org/10.1145/3385412.3386007}
\showDOI{\tempurl}


\bibitem[Calcagno et~al\mbox{.}(2015)]%
        {calcagno2015moving}
\bibfield{author}{\bibinfo{person}{Cristiano Calcagno}, \bibinfo{person}{Dino
  Distefano}, \bibinfo{person}{J{\'e}r{\'e}my Dubreil},
  \bibinfo{person}{Dominik Gabi}, \bibinfo{person}{Pieter Hooimeijer},
  \bibinfo{person}{Martino Luca}, \bibinfo{person}{Peter O’Hearn},
  \bibinfo{person}{Irene Papakonstantinou}, \bibinfo{person}{Jim Purbrick},
  {and} \bibinfo{person}{Dulma Rodriguez}.} \bibinfo{year}{2015}\natexlab{}.
\newblock \showarticletitle{Moving fast with software verification}. In
  \bibinfo{booktitle}{\emph{NASA Formal Methods: 7th International Symposium,
  NFM 2015, Pasadena, CA, USA, April 27-29, 2015, Proceedings 7}}. Springer,
  \bibinfo{pages}{3--11}.
\newblock


\bibitem[Chen et~al\mbox{.}(2023)]%
        {qihongchenSmellyEightEmpirical2023}
\bibfield{author}{\bibinfo{person}{Qihong Chen}, \bibinfo{person}{R{\'u}ben
  C{\^a}mara}, \bibinfo{person}{Jos{\'e} Campos}, \bibinfo{person}{Andr{\'e}
  Souto}, {and} \bibinfo{person}{Iftekhar Ahmed}.}
  \bibinfo{year}{2023}\natexlab{}.
\newblock \showarticletitle{The {{Smelly Eight}}: {{An Empirical Study}} on the
  {{Prevalence}} of {{Code Smells}} in {{Quantum Computing}} - {{Artifact}}}.
\newblock \bibinfo{journal}{\emph{2023 IEEE/ACM 45th International Conference
  on Software Engineering (ICSE)}} (\bibinfo{date}{Jan.} \bibinfo{year}{2023}).
\newblock
\urldef\tempurl%
\url{https://doi.org/10.5281/ZENODO.7556360}
\showDOI{\tempurl}


\bibitem[Dargan(2022)]%
        {darganTopQuantumProgramming2022a}
\bibfield{author}{\bibinfo{person}{James Dargan}.}
  \bibinfo{year}{2022}\natexlab{}.
\newblock \bibinfo{title}{Top 5 {{Quantum Programming Languages}} in 2022}.
\newblock
\newblock


\bibitem[Deutsch and Penrose(1997)]%
        {deutschQuantumComputationalNetworks1997}
\bibfield{author}{\bibinfo{person}{David~Elieser Deutsch} {and}
  \bibinfo{person}{Roger Penrose}.} \bibinfo{year}{1997}\natexlab{}.
\newblock \showarticletitle{Quantum Computational Networks}.
\newblock \bibinfo{journal}{\emph{Proceedings of the Royal Society of London.
  A. Mathematical and Physical Sciences}} \bibinfo{volume}{425},
  \bibinfo{number}{1868} (\bibinfo{date}{Jan.} \bibinfo{year}{1997}),
  \bibinfo{pages}{73--90}.
\newblock
\urldef\tempurl%
\url{https://doi.org/10.1098/rspa.1989.0099}
\showDOI{\tempurl}


\bibitem[Developers(2021a)]%
        {developersCirq2021}
\bibfield{author}{\bibinfo{person}{Cirq Developers}.}
  \bibinfo{year}{2021}\natexlab{a}.
\newblock \bibinfo{title}{Cirq}.
\newblock \bibinfo{howpublished}{Zenodo}.
\newblock
\urldef\tempurl%
\url{https://doi.org/10.5281/zenodo.5182845}
\showDOI{\tempurl}


\bibitem[Developers(2021b)]%
        {qiskit}
\bibfield{author}{\bibinfo{person}{Qiskit Developers}.}
  \bibinfo{year}{2021}\natexlab{b}.
\newblock \bibinfo{title}{Qiskit: {{An}} Open-Source Framework for Quantum
  Computing}.
\newblock
\newblock
\urldef\tempurl%
\url{https://doi.org/10.5281/zenodo.2573505}
\showDOI{\tempurl}


\bibitem[Egretta.Thula(2023)]%
        {egretta.thulaAnswerWhyDoes2023}
\bibfield{author}{\bibinfo{person}{Egretta.Thula}.}
  \bibinfo{year}{2023}\natexlab{}.
\newblock \bibinfo{title}{Answer to "{{Why}} Does Composing a {{Clifford}}
  Circuit to Another Circuit Not Work? ({{Qiskit}})" - {{Quantum Computing
  Stack Exchange}}}.
\newblock
\newblock


\bibitem[Flanagan et~al\mbox{.}(2002)]%
        {flanaganExtendedStaticChecking2002}
\bibfield{author}{\bibinfo{person}{Cormac Flanagan},
  \bibinfo{person}{K.~Rustan~M. Leino}, \bibinfo{person}{Mark Lillibridge},
  \bibinfo{person}{Greg Nelson}, \bibinfo{person}{James~B. Saxe}, {and}
  \bibinfo{person}{Raymie Stata}.} \bibinfo{year}{2002}\natexlab{}.
\newblock \showarticletitle{Extended Static Checking for {{Java}}}.
\newblock \bibinfo{journal}{\emph{ACM SIGPLAN Notices}} \bibinfo{volume}{37},
  \bibinfo{number}{5} (\bibinfo{date}{May} \bibinfo{year}{2002}),
  \bibinfo{pages}{234--245}.
\newblock
\showISSN{0362-1340}
\urldef\tempurl%
\url{https://doi.org/10.1145/543552.512558}
\showDOI{\tempurl}


\bibitem[Fortunato et~al\mbox{.}(2022a)]%
        {fortunatoMutationTestingQuantum2022}
\bibfield{author}{\bibinfo{person}{Daniel Fortunato}, \bibinfo{person}{Jos{\'e}
  Campos}, {and} \bibinfo{person}{Rui Abreu}.}
  \bibinfo{year}{2022}\natexlab{a}.
\newblock \showarticletitle{Mutation Testing of Quantum Programs Written in
  {{QISKit}}}. In \bibinfo{booktitle}{\emph{Proceedings of the {{ACM}}/{{IEEE}}
  44th {{International Conference}} on {{Software Engineering}}: {{Companion
  Proceedings}}}} \emph{(\bibinfo{series}{{{ICSE}} '22})}.
  \bibinfo{publisher}{{Association for Computing Machinery}},
  \bibinfo{address}{{New York, NY, USA}}, \bibinfo{pages}{358--359}.
\newblock
\showISBNx{978-1-4503-9223-5}
\urldef\tempurl%
\url{https://doi.org/10.1145/3510454.3528649}
\showDOI{\tempurl}


\bibitem[Fortunato et~al\mbox{.}(2022b)]%
        {fortunatoQMutPyMutationTesting2022}
\bibfield{author}{\bibinfo{person}{Daniel Fortunato}, \bibinfo{person}{Jos{\'e}
  Campos}, {and} \bibinfo{person}{Rui Abreu}.}
  \bibinfo{year}{2022}\natexlab{b}.
\newblock \showarticletitle{{{QMutPy}}: A Mutation Testing Tool for {{Quantum}}
  Algorithms and Applications in {{Qiskit}}}. In
  \bibinfo{booktitle}{\emph{Proceedings of the 31st {{ACM SIGSOFT International
  Symposium}} on {{Software Testing}} and {{Analysis}}}}
  \emph{(\bibinfo{series}{{{ISSTA}} 2022})}. \bibinfo{publisher}{{Association
  for Computing Machinery}}, \bibinfo{address}{{New York, NY, USA}},
  \bibinfo{pages}{797--800}.
\newblock
\showISBNx{978-1-4503-9379-9}
\urldef\tempurl%
\url{https://doi.org/10.1145/3533767.3543296}
\showDOI{\tempurl}


\bibitem[Ghaleb et~al\mbox{.}(2023)]%
        {ghalebACheckerStaticallyDetecting2023}
\bibfield{author}{\bibinfo{person}{Asem Ghaleb}, \bibinfo{person}{Julia Rubin},
  {and} \bibinfo{person}{Karthik Pattabiraman}.}
  \bibinfo{year}{2023}\natexlab{}.
\newblock \showarticletitle{{{AChecker}}: {{Statically Detecting Smart Contract
  Access Control Vulnerabilities}}}. In \bibinfo{booktitle}{\emph{2023
  {{IEEE}}/{{ACM}} 45th {{International Conference}} on {{Software
  Engineering}} ({{ICSE}})}}. \bibinfo{pages}{945--956}.
\newblock
\showISSN{1558-1225}
\urldef\tempurl%
\url{https://doi.org/10.1109/ICSE48619.2023.00087}
\showDOI{\tempurl}


\bibitem[Google(2015)]%
        {ErrorProne}
\bibfield{author}{\bibinfo{person}{Google}.} \bibinfo{year}{2015}\natexlab{}.
\newblock \bibinfo{title}{Error Prone: static analysis tool for Java}.
\newblock \bibinfo{howpublished}{\url{http://errorprone.info/}}.
\newblock


\bibitem[Habib and Pradel(2018)]%
        {habibHowManyAll2018}
\bibfield{author}{\bibinfo{person}{Andrew Habib} {and} \bibinfo{person}{Michael
  Pradel}.} \bibinfo{year}{2018}\natexlab{}.
\newblock \showarticletitle{How {{Many}} of {{All Bugs Do We Find}}? {{A
  Study}} of {{Static Bug Detectors}}}. In \bibinfo{booktitle}{\emph{2018 33rd
  {{IEEE}}/{{ACM International Conference}} on {{Automated Software
  Engineering}} ({{ASE}})}}. \bibinfo{pages}{317--328}.
\newblock
\showISSN{2643-1572}
\urldef\tempurl%
\url{https://doi.org/10.1145/3238147.3238213}
\showDOI{\tempurl}


\bibitem[Huang and Martonosi(2019)]%
        {huangStatisticalAssertionsValidating2019}
\bibfield{author}{\bibinfo{person}{Yipeng Huang} {and}
  \bibinfo{person}{Margaret Martonosi}.} \bibinfo{year}{2019}\natexlab{}.
\newblock \showarticletitle{Statistical Assertions for Validating Patterns and
  Finding Bugs in Quantum Programs}. In \bibinfo{booktitle}{\emph{Proceedings
  of the 46th {{International Symposium}} on {{Computer Architecture}}}}
  \emph{(\bibinfo{series}{{{ISCA}} '19})}. \bibinfo{publisher}{{Association for
  Computing Machinery}}, \bibinfo{address}{{New York, NY, USA}},
  \bibinfo{pages}{541--553}.
\newblock
\showISBNx{978-1-4503-6669-4}
\urldef\tempurl%
\url{https://doi.org/10.1145/3307650.3322213}
\showDOI{\tempurl}


\bibitem[JavadiAbhari et~al\mbox{.}(2014)]%
        {javadiabhariScaffCCFrameworkCompilation2014}
\bibfield{author}{\bibinfo{person}{Ali JavadiAbhari}, \bibinfo{person}{Shruti
  Patil}, \bibinfo{person}{Daniel Kudrow}, \bibinfo{person}{Jeff Heckey},
  \bibinfo{person}{Alexey Lvov}, \bibinfo{person}{Frederic~T. Chong}, {and}
  \bibinfo{person}{Margaret Martonosi}.} \bibinfo{year}{2014}\natexlab{}.
\newblock \showarticletitle{{{ScaffCC}}: A Framework for Compilation and
  Analysis of Quantum Computing Programs}. In
  \bibinfo{booktitle}{\emph{Proceedings of the 11th {{ACM Conference}} on
  {{Computing Frontiers}}}} \emph{(\bibinfo{series}{{{CF}} '14})}.
  \bibinfo{publisher}{{Association for Computing Machinery}},
  \bibinfo{address}{{New York, NY, USA}}, \bibinfo{pages}{1--10}.
\newblock
\showISBNx{978-1-4503-2870-8}
\urldef\tempurl%
\url{https://doi.org/10.1145/2597917.2597939}
\showDOI{\tempurl}


\bibitem[Johnson et~al\mbox{.}(2013)]%
        {johnsonWhyDonSoftware2013}
\bibfield{author}{\bibinfo{person}{Brittany Johnson}, \bibinfo{person}{Yoonki
  Song}, \bibinfo{person}{Emerson {Murphy-Hill}}, {and} \bibinfo{person}{Robert
  Bowdidge}.} \bibinfo{year}{2013}\natexlab{}.
\newblock \showarticletitle{Why Don't Software Developers Use Static Analysis
  Tools to Find Bugs?}. In \bibinfo{booktitle}{\emph{Proceedings of the 2013
  {{International Conference}} on {{Software Engineering}}}}
  \emph{(\bibinfo{series}{{{ICSE}} '13})}. \bibinfo{publisher}{{IEEE Press}},
  \bibinfo{address}{{San Francisco, CA, USA}}, \bibinfo{pages}{672--681}.
\newblock
\showISBNx{978-1-4673-3076-3}


\bibitem[Just et~al\mbox{.}(2014)]%
        {justDefects4JDatabaseExisting2014}
\bibfield{author}{\bibinfo{person}{Ren{\'e} Just}, \bibinfo{person}{Darioush
  Jalali}, {and} \bibinfo{person}{Michael~D. Ernst}.}
  \bibinfo{year}{2014}\natexlab{}.
\newblock \showarticletitle{{{Defects4J}}: A Database of Existing Faults to
  Enable Controlled Testing Studies for {{Java}} Programs}. In
  \bibinfo{booktitle}{\emph{Proceedings of the 2014 {{International Symposium}}
  on {{Software Testing}} and {{Analysis}}}} \emph{(\bibinfo{series}{{{ISSTA}}
  2014})}. \bibinfo{publisher}{{Association for Computing Machinery}},
  \bibinfo{address}{{New York, NY, USA}}, \bibinfo{pages}{437--440}.
\newblock
\showISBNx{978-1-4503-2645-2}
\urldef\tempurl%
\url{https://doi.org/10.1145/2610384.2628055}
\showDOI{\tempurl}


\bibitem[Kaul et~al\mbox{.}(2023)]%
        {kaulUniformRepresentationClassical2023}
\bibfield{author}{\bibinfo{person}{Maximilian Kaul}, \bibinfo{person}{Alexander
  K{\"u}chler}, {and} \bibinfo{person}{Christian Banse}.}
  \bibinfo{year}{2023}\natexlab{}.
\newblock \bibinfo{title}{A {{Uniform Representation}} of {{Classical}} and
  {{Quantum Source Code}} for {{Static Code Analysis}}}.
\newblock
\newblock
\urldef\tempurl%
\url{https://doi.org/10.48550/arXiv.2308.06113}
\showDOI{\tempurl}
\showeprint[arxiv]{2308.06113}~[cs]


\bibitem[Lagouvardos et~al\mbox{.}(2020)]%
        {lagouvardosStaticAnalysisShape2020}
\bibfield{author}{\bibinfo{person}{Sifis Lagouvardos}, \bibinfo{person}{Julian
  Dolby}, \bibinfo{person}{Neville Grech}, \bibinfo{person}{Anastasios
  Antoniadis}, {and} \bibinfo{person}{Yannis Smaragdakis}.}
  \bibinfo{year}{2020}\natexlab{}.
\newblock \showarticletitle{Static {{Analysis}} of {{Shape}} in {{TensorFlow
  Programs}}}. In \bibinfo{booktitle}{\emph{34th {{European Conference}} on
  {{Object-Oriented Programming}} ({{ECOOP}} 2020)}}
  \emph{(\bibinfo{series}{Leibniz {{International Proceedings}} in
  {{Informatics}} ({{LIPIcs}})}, Vol.~\bibinfo{volume}{166})},
  \bibfield{editor}{\bibinfo{person}{Robert Hirschfeld} {and}
  \bibinfo{person}{Tobias Pape}} (Eds.). \bibinfo{publisher}{{Schloss
  Dagstuhl\textendash Leibniz-Zentrum f\"ur Informatik}},
  \bibinfo{address}{{Dagstuhl, Germany}}, \bibinfo{pages}{15:1--15:29}.
\newblock
\showISBNx{978-3-95977-154-2}
\showISSN{1868-8969}
\urldef\tempurl%
\url{https://doi.org/10.4230/LIPIcs.ECOOP.2020.15}
\showDOI{\tempurl}


\bibitem[Li et~al\mbox{.}(2023)]%
        {liQASMBenchLowLevelQuantum2023}
\bibfield{author}{\bibinfo{person}{Ang Li}, \bibinfo{person}{Samuel Stein},
  \bibinfo{person}{Sriram Krishnamoorthy}, {and} \bibinfo{person}{James Ang}.}
  \bibinfo{year}{2023}\natexlab{}.
\newblock \showarticletitle{{{QASMBench}}: {{A Low-Level Quantum Benchmark
  Suite}} for {{NISQ Evaluation}} and {{Simulation}}}.
\newblock \bibinfo{journal}{\emph{ACM Transactions on Quantum Computing}}
  \bibinfo{volume}{4}, \bibinfo{number}{2} (\bibinfo{date}{Feb.}
  \bibinfo{year}{2023}), \bibinfo{pages}{10:1--10:26}.
\newblock
\urldef\tempurl%
\url{https://doi.org/10.1145/3550488}
\showDOI{\tempurl}


\bibitem[Li et~al\mbox{.}(2020)]%
        {liProjectionbasedRuntimeAssertions2020}
\bibfield{author}{\bibinfo{person}{Gushu Li}, \bibinfo{person}{Li Zhou},
  \bibinfo{person}{Nengkun Yu}, \bibinfo{person}{Yufei Ding},
  \bibinfo{person}{Mingsheng Ying}, {and} \bibinfo{person}{Yuan Xie}.}
  \bibinfo{year}{2020}\natexlab{}.
\newblock \showarticletitle{Projection-Based Runtime Assertions for Testing and
  Debugging {{Quantum}} Programs}.
\newblock \bibinfo{journal}{\emph{Proceedings of the ACM on Programming
  Languages}} \bibinfo{volume}{4}, \bibinfo{number}{OOPSLA}
  (\bibinfo{date}{Nov.} \bibinfo{year}{2020}), \bibinfo{pages}{150:1--150:29}.
\newblock
\urldef\tempurl%
\url{https://doi.org/10.1145/3428218}
\showDOI{\tempurl}


\bibitem[Livshits et~al\mbox{.}(2015)]%
        {Livshits2015}
\bibfield{author}{\bibinfo{person}{Benjamin Livshits}, \bibinfo{person}{Manu
  Sridharan}, \bibinfo{person}{Yannis Smaragdakis}, \bibinfo{person}{Ondrej
  Lhot{\'{a}}k}, \bibinfo{person}{Jos{\'{e}}~Nelson Amaral},
  \bibinfo{person}{Bor{-}Yuh~Evan Chang}, \bibinfo{person}{Samuel~Z. Guyer},
  \bibinfo{person}{Uday~P. Khedker}, \bibinfo{person}{Anders M{\o}ller}, {and}
  \bibinfo{person}{Dimitrios Vardoulakis}.} \bibinfo{year}{2015}\natexlab{}.
\newblock \showarticletitle{In defense of soundiness: a manifesto}.
\newblock \bibinfo{journal}{\emph{Commun. {ACM}}} \bibinfo{volume}{58},
  \bibinfo{number}{2} (\bibinfo{year}{2015}), \bibinfo{pages}{44--46}.
\newblock


\bibitem[Long and Zhao(2023)]%
        {longEquivalenceIdentityUnitarity2023}
\bibfield{author}{\bibinfo{person}{Peixun Long} {and} \bibinfo{person}{Jianjun
  Zhao}.} \bibinfo{year}{2023}\natexlab{}.
\newblock \bibinfo{title}{Equivalence, {{Identity}}, and {{Unitarity Checking}}
  in {{Black-Box Testing}} of {{Quantum Programs}}}.
\newblock
\newblock
\urldef\tempurl%
\url{https://doi.org/10.48550/arXiv.2307.01481}
\showDOI{\tempurl}
\showeprint[arxiv]{2307.01481}~[quant-ph]


\bibitem[Luo et~al\mbox{.}(2022)]%
        {luoComprehensiveStudyBug2022a}
\bibfield{author}{\bibinfo{person}{Junjie Luo}, \bibinfo{person}{Pengzhan
  Zhao}, \bibinfo{person}{Zhongtao Miao}, \bibinfo{person}{Shuhan Lan}, {and}
  \bibinfo{person}{Jianjun Zhao}.} \bibinfo{year}{2022}\natexlab{}.
\newblock \showarticletitle{A {{Comprehensive Study}} of {{Bug Fixes}} in
  {{Quantum Programs}}}. In \bibinfo{booktitle}{\emph{2022 {{IEEE International
  Conference}} on {{Software Analysis}}, {{Evolution}} and {{Reengineering}}
  ({{SANER}})}}. \bibinfo{pages}{1239--1246}.
\newblock
\showISSN{1534-5351}
\urldef\tempurl%
\url{https://doi.org/10.1109/SANER53432.2022.00147}
\showDOI{\tempurl}


\bibitem[Miranskyy et~al\mbox{.}(2020)]%
        {miranskyyYourQuantumProgram2020}
\bibfield{author}{\bibinfo{person}{Andriy Miranskyy}, \bibinfo{person}{Lei
  Zhang}, {and} \bibinfo{person}{Javad Doliskani}.}
  \bibinfo{year}{2020}\natexlab{}.
\newblock \showarticletitle{Is {{Your Quantum Program Bug-Free}}?}
\newblock \bibinfo{journal}{\emph{Proceedings of the ACM/IEEE 42nd
  International Conference on Software Engineering: New Ideas and Emerging
  Results}} (\bibinfo{date}{June} \bibinfo{year}{2020}),
  \bibinfo{pages}{29--32}.
\newblock
\urldef\tempurl%
\url{https://doi.org/10.1145/3377816.3381731}
\showDOI{\tempurl}
\showeprint[arxiv]{2001.10870}


\bibitem[Nielsen et~al\mbox{.}(2002)]%
        {nielsenQuantumComputationQuantum2002}
\bibfield{author}{\bibinfo{person}{Michael~A Nielsen}, \bibinfo{person}{Isaac
  Chuang}, {and} \bibinfo{person}{Lov~K Grover}.}
  \bibinfo{year}{2002}\natexlab{}.
\newblock \showarticletitle{Quantum {{Computation}} and {{Quantum
  Information}}}.
\newblock \bibinfo{journal}{\emph{Am. J. Phys.}} \bibinfo{volume}{70},
  \bibinfo{number}{5} (\bibinfo{year}{2002}), \bibinfo{pages}{4}.
\newblock


\bibitem[Paltenghi and Pradel(2022)]%
        {paltenghiBugsQuantumComputing2022}
\bibfield{author}{\bibinfo{person}{Matteo Paltenghi} {and}
  \bibinfo{person}{Michael Pradel}.} \bibinfo{year}{2022}\natexlab{}.
\newblock \showarticletitle{Bugs in {{Quantum}} Computing Platforms: An
  Empirical Study}.
\newblock \bibinfo{journal}{\emph{Proceedings of the ACM on Programming
  Languages}} \bibinfo{volume}{6}, \bibinfo{number}{OOPSLA1}
  (\bibinfo{date}{April} \bibinfo{year}{2022}), \bibinfo{pages}{86:1--86:27}.
\newblock
\urldef\tempurl%
\url{https://doi.org/10.1145/3527330}
\showDOI{\tempurl}


\bibitem[Paltenghi and Pradel(2023)]%
        {paltenghiMorphQMetamorphicTesting2023}
\bibfield{author}{\bibinfo{person}{Matteo Paltenghi} {and}
  \bibinfo{person}{Michael Pradel}.} \bibinfo{year}{2023}\natexlab{}.
\newblock \showarticletitle{{{MorphQ}}: {{Metamorphic Testing}} of the {{Qiskit
  Quantum Computing Platform}}}. In \bibinfo{booktitle}{\emph{Proceedings of
  the 45th {{International Conference}} on {{Software Engineering}}}}
  \emph{(\bibinfo{series}{{{ICSE}} '23})}. \bibinfo{publisher}{{IEEE Press}},
  \bibinfo{address}{{Melbourne, Victoria, Australia}},
  \bibinfo{pages}{2413--2424}.
\newblock
\showISBNx{978-1-66545-701-9}
\urldef\tempurl%
\url{https://doi.org/10.1109/ICSE48619.2023.00202}
\showDOI{\tempurl}


\bibitem[Paradis et~al\mbox{.}(2021)]%
        {paradisUnqompSynthesizingUncomputation2021}
\bibfield{author}{\bibinfo{person}{Anouk Paradis}, \bibinfo{person}{Benjamin
  Bichsel}, \bibinfo{person}{Samuel Steffen}, {and} \bibinfo{person}{Martin
  Vechev}.} \bibinfo{year}{2021}\natexlab{}.
\newblock \showarticletitle{Unqomp: Synthesizing Uncomputation in {{Quantum}}
  Circuits}. In \bibinfo{booktitle}{\emph{Proceedings of the 42nd {{ACM SIGPLAN
  International Conference}} on {{Programming Language Design}} and
  {{Implementation}}}} \emph{(\bibinfo{series}{{{PLDI}} 2021})}.
  \bibinfo{publisher}{{Association for Computing Machinery}},
  \bibinfo{address}{{New York, NY, USA}}, \bibinfo{pages}{222--236}.
\newblock
\showISBNx{978-1-4503-8391-2}
\urldef\tempurl%
\url{https://doi.org/10.1145/3453483.3454040}
\showDOI{\tempurl}


\bibitem[Perdrix(2008)]%
        {perdrixQuantumEntanglementAnalysis2008}
\bibfield{author}{\bibinfo{person}{Simon Perdrix}.}
  \bibinfo{year}{2008}\natexlab{}.
\newblock \showarticletitle{Quantum {{Entanglement Analysis Based}} on
  {{Abstract Interpretation}}}. In \bibinfo{booktitle}{\emph{Static
  {{Analysis}}}} \emph{(\bibinfo{series}{Lecture {{Notes}} in {{Computer
  Science}}})}, \bibfield{editor}{\bibinfo{person}{Mar{\'i}a Alpuente} {and}
  \bibinfo{person}{Germ{\'a}n Vidal}} (Eds.). \bibinfo{publisher}{{Springer}},
  \bibinfo{address}{{Berlin, Heidelberg}}, \bibinfo{pages}{270--282}.
\newblock
\showISBNx{978-3-540-69166-2}
\urldef\tempurl%
\url{https://doi.org/10.1007/978-3-540-69166-2_18}
\showDOI{\tempurl}


\bibitem[Sivarajah et~al\mbox{.}(2020)]%
        {sivarajahKetRetargetableCompiler2020}
\bibfield{author}{\bibinfo{person}{Seyon Sivarajah}, \bibinfo{person}{Silas
  Dilkes}, \bibinfo{person}{Alexander Cowtan}, \bibinfo{person}{Will Simmons},
  \bibinfo{person}{Alec Edgington}, {and} \bibinfo{person}{Ross Duncan}.}
  \bibinfo{year}{2020}\natexlab{}.
\newblock \showarticletitle{T|ket{$\rangle$}: A Retargetable Compiler for
  {{NISQ}} Devices}.
\newblock \bibinfo{journal}{\emph{Quantum Science and Technology}}
  \bibinfo{volume}{6}, \bibinfo{number}{1} (\bibinfo{date}{Nov.}
  \bibinfo{year}{2020}), \bibinfo{pages}{014003}.
\newblock
\showISSN{2058-9565}
\urldef\tempurl%
\url{https://doi.org/10.1088/2058-9565/ab8e92}
\showDOI{\tempurl}


\bibitem[Svore et~al\mbox{.}(2018)]%
        {svoreEnablingScalableQuantum2018}
\bibfield{author}{\bibinfo{person}{Krysta Svore}, \bibinfo{person}{Alan
  Geller}, \bibinfo{person}{Matthias Troyer}, \bibinfo{person}{John Azariah},
  \bibinfo{person}{Christopher Granade}, \bibinfo{person}{Bettina Heim},
  \bibinfo{person}{Vadym Kliuchnikov}, \bibinfo{person}{Mariia Mykhailova},
  \bibinfo{person}{Andres Paz}, {and} \bibinfo{person}{Martin Roetteler}.}
  \bibinfo{year}{2018}\natexlab{}.
\newblock \showarticletitle{Q\#: {{Enabling Scalable Quantum Computing}} and
  {{Development}} with a {{High-level DSL}}}. In
  \bibinfo{booktitle}{\emph{Proceedings of the {{Real World Domain Specific
  Languages Workshop}} 2018}} \emph{(\bibinfo{series}{{{RWDSL2018}}})}.
  \bibinfo{publisher}{{Association for Computing Machinery}},
  \bibinfo{address}{{New York, NY, USA}}, \bibinfo{pages}{1--10}.
\newblock
\showISBNx{978-1-4503-6355-6}
\urldef\tempurl%
\url{https://doi.org/10.1145/3183895.3183901}
\showDOI{\tempurl}


\bibitem[{user19571}(2022)]%
        {user19571QuestionRemoveInactive2022}
\bibfield{author}{\bibinfo{person}{{user19571}}.}
  \bibinfo{year}{2022}\natexlab{}.
\newblock \bibinfo{title}{Question: "{{Remove Inactive Qubits}} from {{Qiskit
  Circuit}}" - {{Quantum Computing Stack Exchange}}}.
\newblock
\newblock


\bibitem[Wang et~al\mbox{.}(2021a)]%
        {wangPosterFuzzTesting2021}
\bibfield{author}{\bibinfo{person}{Jiyuan Wang}, \bibinfo{person}{Fucheng Ma},
  {and} \bibinfo{person}{Yu Jiang}.} \bibinfo{year}{2021}\natexlab{a}.
\newblock \showarticletitle{Poster: {{Fuzz Testing}} of {{Quantum Program}}}.
  In \bibinfo{booktitle}{\emph{2021 14th {{IEEE Conference}} on {{Software
  Testing}}, {{Verification}} and {{Validation}} ({{ICST}})}}.
  \bibinfo{pages}{466--469}.
\newblock
\showISSN{2159-4848}
\urldef\tempurl%
\url{https://doi.org/10.1109/ICST49551.2021.00061}
\showDOI{\tempurl}


\bibitem[Wang et~al\mbox{.}(2021b)]%
        {wangQDiffDifferentialTesting2021}
\bibfield{author}{\bibinfo{person}{Jiyuan Wang}, \bibinfo{person}{Qian Zhang},
  \bibinfo{person}{Guoqing~Harry Xu}, {and} \bibinfo{person}{Miryung Kim}.}
  \bibinfo{year}{2021}\natexlab{b}.
\newblock \showarticletitle{{{QDiff}}: {{Differential Testing}} of {{Quantum
  Software Stacks}}}. In \bibinfo{booktitle}{\emph{2021 36th {{IEEE}}/{{ACM
  International Conference}} on {{Automated Software Engineering}} ({{ASE}})}}.
  \bibinfo{pages}{692--704}.
\newblock
\showISSN{2643-1572}
\urldef\tempurl%
\url{https://doi.org/10.1109/ASE51524.2021.9678792}
\showDOI{\tempurl}


\bibitem[Weigold et~al\mbox{.}(2021)]%
        {weigoldEncodingPatternsQuantum2021}
\bibfield{author}{\bibinfo{person}{Manuela Weigold}, \bibinfo{person}{Johanna
  Barzen}, \bibinfo{person}{Frank Leymann}, {and} \bibinfo{person}{Marie
  Salm}.} \bibinfo{year}{2021}\natexlab{}.
\newblock \showarticletitle{Encoding Patterns for Quantum Algorithms}.
\newblock \bibinfo{journal}{\emph{IET Quantum Communication}}
  \bibinfo{volume}{2}, \bibinfo{number}{4} (\bibinfo{year}{2021}),
  \bibinfo{pages}{141--152}.
\newblock
\showISSN{2632-8925}
\urldef\tempurl%
\url{https://doi.org/10.1049/qtc2.12032}
\showDOI{\tempurl}


\bibitem[Xia et~al\mbox{.}(2024)]%
        {xiaFuzz4AllUniversalFuzzing2024}
\bibfield{author}{\bibinfo{person}{Chunqiu~Steven Xia}, \bibinfo{person}{Matteo
  Paltenghi}, \bibinfo{person}{Jia Le~Tian}, \bibinfo{person}{Michael Pradel},
  {and} \bibinfo{person}{Lingming Zhang}.} \bibinfo{year}{2024}\natexlab{}.
\newblock \showarticletitle{{{Fuzz4All}}: {{Universal Fuzzing}} with {{Large
  Language Models}}}. In \bibinfo{booktitle}{\emph{Proceedings of the
  {{IEEE}}/{{ACM}} 46th {{International Conference}} on {{Software
  Engineering}}}} \emph{(\bibinfo{series}{{{ICSE}} '24})}.
  \bibinfo{publisher}{{Association for Computing Machinery}},
  \bibinfo{address}{{New York, NY, USA}}, \bibinfo{pages}{1--13}.
\newblock
\showISBNx{9798400702174}
\urldef\tempurl%
\url{https://doi.org/10.1145/3597503.3639121}
\showDOI{\tempurl}


\bibitem[Xia and Zhao(2023)]%
        {xiaStaticEntanglementAnalysis2023}
\bibfield{author}{\bibinfo{person}{Shangzhou Xia} {and}
  \bibinfo{person}{Jianjun Zhao}.} \bibinfo{year}{2023}\natexlab{}.
\newblock \showarticletitle{Static {{Entanglement Analysis}} of {{Quantum
  Programs}}}. In \bibinfo{booktitle}{\emph{2023 {{IEEE}}/{{ACM}} 4th
  {{International Workshop}} on {{Quantum Software Engineering}} ({{Q-SE}})}}.
  \bibinfo{pages}{42--49}.
\newblock
\urldef\tempurl%
\url{https://doi.org/10.1109/Q-SE59154.2023.00013}
\showDOI{\tempurl}


\bibitem[Yamaguchi et~al\mbox{.}(2014)]%
        {yamaguchiModelingDiscoveringVulnerabilities2014}
\bibfield{author}{\bibinfo{person}{Fabian Yamaguchi}, \bibinfo{person}{Nico
  Golde}, \bibinfo{person}{Daniel Arp}, {and} \bibinfo{person}{Konrad Rieck}.}
  \bibinfo{year}{2014}\natexlab{}.
\newblock \showarticletitle{Modeling and {{Discovering Vulnerabilities}} with
  {{Code Property Graphs}}}. In \bibinfo{booktitle}{\emph{2014 {{IEEE
  Symposium}} on {{Security}} and {{Privacy}}}}. \bibinfo{pages}{590--604}.
\newblock
\showISSN{2375-1207}
\urldef\tempurl%
\url{https://doi.org/10.1109/SP.2014.44}
\showDOI{\tempurl}


\bibitem[Yu and Palsberg(2021)]%
        {yuQuantumAbstractInterpretation2021}
\bibfield{author}{\bibinfo{person}{Nengkun Yu} {and} \bibinfo{person}{Jens
  Palsberg}.} \bibinfo{year}{2021}\natexlab{}.
\newblock \showarticletitle{Quantum Abstract Interpretation}. In
  \bibinfo{booktitle}{\emph{Proceedings of the 42nd {{ACM SIGPLAN International
  Conference}} on {{Programming Language Design}} and {{Implementation}}}}
  \emph{(\bibinfo{series}{{{PLDI}} 2021})}. \bibinfo{publisher}{{Association
  for Computing Machinery}}, \bibinfo{address}{{New York, NY, USA}},
  \bibinfo{pages}{542--558}.
\newblock
\showISBNx{978-1-4503-8391-2}
\urldef\tempurl%
\url{https://doi.org/10.1145/3453483.3454061}
\showDOI{\tempurl}


\bibitem[Zhao et~al\mbox{.}(2023a)]%
        {zhaoBugs4QBenchmarkExisting2023}
\bibfield{author}{\bibinfo{person}{Pengzhan Zhao}, \bibinfo{person}{Zhongtao
  Miao}, \bibinfo{person}{Shuhan Lan}, {and} \bibinfo{person}{Jianjun Zhao}.}
  \bibinfo{year}{2023}\natexlab{a}.
\newblock \showarticletitle{{{Bugs4Q}}: {{A}} Benchmark of Existing Bugs to
  Enable Controlled Testing and Debugging Studies for Quantum Programs}.
\newblock \bibinfo{journal}{\emph{Journal of Systems and Software}}
  \bibinfo{volume}{205} (\bibinfo{date}{Nov.} \bibinfo{year}{2023}),
  \bibinfo{pages}{111805}.
\newblock
\showISSN{0164-1212}
\urldef\tempurl%
\url{https://doi.org/10.1016/j.jss.2023.111805}
\showDOI{\tempurl}


\bibitem[Zhao et~al\mbox{.}(2023b)]%
        {zhaoQCheckerDetectingBugs2023}
\bibfield{author}{\bibinfo{person}{Pengzhan Zhao}, \bibinfo{person}{Xiongfei
  Wu}, \bibinfo{person}{Zhuo Li}, {and} \bibinfo{person}{Jianjun Zhao}.}
  \bibinfo{year}{2023}\natexlab{b}.
\newblock \bibinfo{title}{{{QChecker}}: {{Detecting Bugs}} in {{Quantum
  Programs}} via {{Static Analysis}}}.
\newblock
\newblock
\urldef\tempurl%
\url{https://doi.org/10.48550/arXiv.2304.04387}
\showDOI{\tempurl}
\showeprint[arxiv]{2304.04387}~[cs]


\bibitem[Zhao et~al\mbox{.}(2021a)]%
        {zhaoIdentifyingBugPatterns2021}
\bibfield{author}{\bibinfo{person}{Pengzhan Zhao}, \bibinfo{person}{Jianjun
  Zhao}, {and} \bibinfo{person}{Lei Ma}.} \bibinfo{year}{2021}\natexlab{a}.
\newblock \showarticletitle{Identifying {{Bug Patterns}} in {{Quantum
  Programs}}}. In \bibinfo{booktitle}{\emph{2021 {{IEEE}}/{{ACM}} 2nd
  {{International Workshop}} on {{Quantum Software Engineering}} ({{Q-SE}})}}.
  \bibinfo{pages}{16--21}.
\newblock
\showISBNx{978-1-66544-462-0}
\urldef\tempurl%
\url{https://doi.org/10.1109/Q-SE52541.2021.00011}
\showDOI{\tempurl}


\bibitem[Zhao et~al\mbox{.}(2021b)]%
        {zhaoBugs4QBenchmarkReal2021}
\bibfield{author}{\bibinfo{person}{Pengzhan Zhao}, \bibinfo{person}{Jianjun
  Zhao}, \bibinfo{person}{Zhongtao Miao}, {and} \bibinfo{person}{Shuhan Lan}.}
  \bibinfo{year}{2021}\natexlab{b}.
\newblock \showarticletitle{{{Bugs4Q}}: {{A Benchmark}} of {{Real Bugs}} for
  {{Quantum Programs}}}. In \bibinfo{booktitle}{\emph{2021 36th {{IEEE}}/{{ACM
  International Conference}} on {{Automated Software Engineering}} ({{ASE}})}}.
\newblock
\showISSN{2643-1572}
\urldef\tempurl%
\url{https://doi.org/10.1109/ASE51524.2021.9678908}
\showDOI{\tempurl}


\end{thebibliography}

\end{document}